\documentclass[twocolumn,epjc3]{svjour3}

\RequirePackage[T1]{fontenc}

\smartqed  % flush right qed marks, e.g. at end of proof

\RequirePackage{graphicx}
\RequirePackage{mathptmx}      % use Times fonts if available on your TeX system
\RequirePackage{flushend}
\RequirePackage[numbers,sort&compress]{natbib}
\RequirePackage[colorlinks,citecolor=blue,urlcolor=blue,linkcolor=blue]{hyperref}
\usepackage{subfigure}
\usepackage{cuted}

\journalname{Eur. Phys. J. C}

\begin{document}
\title{Phase Diagrams of Charged Compact Boson Stars} 

\author{Sanjeev Kumar\thanksref{e1,adr1,adr2}\and Usha Kulshreshtha\thanksref{e2,adr3,adr4}\and Daya Shankar Kulshreshtha\thanksref{e3,adr1,adr4}\and Jutta Kunz\thanksref{e4,adr4}}
%\email{sanjeev.kumar.ka@gmail.com}

\institute{Department of Physics and Astrophysics, University of Delhi, Delhi-110007, India\label{adr1}\and Department of Physics, Pt. NRS Govt. College, M.D. University, Rohtak-124001, India\label{adr2} \and Department of Physics, Kirori Mal College, University of Delhi, Delhi-110007, India\label{adr3}\and  Institut f\"ur Physik, Universit\"at Oldenburg, Postfach 2503, D-26111 Oldenburg, Germany\label{adr4}}

\thankstext{e1}{e-mail: sanjeev.kumar.ka@gmail.com}
\thankstext{e2}{e-mail: ushakulsh@gmail.com}
\thankstext{e3}{e-mail: dskulsh@gmail.com}
\thankstext{e4}{e-mail: jutta.kunz@uni-oldenburg.de}

\date{Received: date / Accepted: date}
\maketitle

\begin{abstract}
Compact boson stars, whose scalar field vanishes identically 
in the exterior region, arise in a theory involving a {\it massless} 
complex scalar field with a conical potential, when coupled to gravity.
Their charged compact generalizations, obtained in the presence of a 
U(1) gauge field, exhibit further interesting features.
On the one hand, charged compact boson shells can arise, whose scalar field 
vanishes also in the central region,
while on the other hand, the domain of existence of charged compact boson stars
exhibits bifurcation points. First 2D phase diagrams have been studied before.
Here we extend these earlier studies to a larger range of the variables
and study additional phase diagrams. We then extend these studies 
to obtain 3D phase diagrams and present these with a detailed discussion 
of their various regions with respect to the bifurcation points
and argue, that there is an infinite series of such bifurcation points.
Thus the theory is seen to contain rich physics in a particular domain 
of the phase diagrams. We also discuss the dependence of the fields 
on the dimensionless radial coordinate for some representative points 
of the phase trajectories in the phase diagrams of the theory.
\end{abstract}

\section{Introduction}\label{sec:int}

Boson stars represent localized solutions of scalar fields coupled to gravity
(see the reviews \cite{Lee:1991ax,Jetzer:1991jr,Mielke:2000mh,Schunck:2003kk,Liebling:2012fv}).
In their original form, they are based on a complex scalar field 
with a mass term only \cite{Kaup:1968zz,Feinblum:1968nwc,Ruffini:1969qy},
while later work has included self-interactions of the scalar field,
allowing for much higher masses of the resulting boson stars
%for a given boson mass
\cite{Colpi:1986ye,Friedberg:1986tq}.

Since being first conceived half a century ago,
boson stars have received tremendous interest.
Much of this interest resides in their potential astrophysical
applications. These range from being promising dark matter candidates
all the way to representing highly compact supermassive objects 
in galactic centers \cite{Lee:1991ax,Jetzer:1991jr,Mielke:2000mh,Schunck:2003kk,Liebling:2012fv}).
Among the many applications 
accretion disks around boson stars have been considered
\cite{Meliani:2015zta,Meliani:2017ktw}
or the gravitational-wave signatures of boson stars
and binary systems of boson stars
have been studied \cite{Macedo:2013jja,Palenzuela:2017kcg}.
In addition, boson stars have found applications based on the AdS/CFT
conjecture \cite{Hartmann:2012wa}.

However, boson stars are also interesting from a more theoretical point of view.
On the one hand, they allow a study of numerous features of more complicated
systems, and thus represent a nice model to learn and analyze.
On the other hand, they contain the freedom to allow for new fascinating
properties, not encountered before -- at least not under similar 
circumstances. One of the most impressive recent examples in this context
represents the discovery of hairy black holes based on boson stars
\cite{Herdeiro:2014goa,Herdeiro:2015tia}.
Another recent surprise was the realization that rotating boson stars
allow for static orbits \cite{Collodel:2017end}.

When considering the properties of boson stars, 
first of all their mass, their particle number and their size,
are of interest. The particle number of boson stars
arises from the U(1) symmetry of the underlying theory,
which gives rise to a conserved current and thus an associated charge.
When the U(1) symmetry is made a local symmetry, charged boson stars
arise \cite{Jetzer:1989av,Jetzer:1989us,Jetzer:1993nk,Pugliese:2013gsa,Delgado:2016jxq,Collodel:2019ohy}.
While their particle number is proportional to their charge, 
their basic features seem to only straightforwardly generalize those of 
their uncharged brethren.
This changes fundamentally, when a different type of scalar field potential
is employed, namely a conical potential
\cite{Arodz:2008jk,Arodz:2008nm}.

First of all, a conical potential allows for compact boson star solutions,
where \textit{compact} is not meant to describe the physical property of
having a large mass residing in an area with a small radius, 
as typically employed in connection with neutron stars or black holes.
Here \textit{compact} is meant to describe a solution whose scalar field
vanishes outside a compact region of space identically.
Thus while ordinary boson stars do not feature sharp boundaries, since
their scalar field exhibits only an exponential fall-off,
in compact boson stars, the scalar field is completely confined
to a finite region of space 
\cite{Kleihaus:2009kr,Kleihaus:2010ep,Hartmann:2012da,Hartmann:2013kna,Kumar:2014kna,Kumar:2015sia,Kumar:2016oop,Kumar:2016sxx,Kumar:2017zms}.

In the presence of a gauge field, however, charged compact boson stars exhibit
new exciting features. One of these corresponds to the fact, that 
charged compact boson shells arise \cite{Kleihaus:2009kr,Kleihaus:2010ep}.
Here the boson field is finite only inside a shell and vanishes identically
outside this shell. Interestingly, there does not even have to be a
vacuum in the central region of the spacetime. Instead, a black hole may
reside there, that will then be surrounded by a matter shell
as discussed in detail before \cite{Kleihaus:2010ep}.
Another interesting feature of these charged compact boson stars represents 
the presence of bifurcations points in the domain of existence
\cite{Kleihaus:2009kr,Kumar:2017zms}.

Studies of the charged compact boson stars in a theory involving a {\it massive} complex scalar field with a conical potential coupled to a U(1) gauge field and gravity were undertaken in Refs.~\cite{Kumar:2014kna,Kumar:2015sia} and \cite{Kumar:2016oop,Kumar:2016sxx}. In this work,  the 2D and 3D phase diagrams were investigated in rather great detail, and multiple bifurcation points showing the presence of rather rich physics were obtained for a particular range of the parameters of the theory. These studies were undertaken in the absence \cite{Kumar:2014kna,Kumar:2015sia} as well as in the presence \cite{Kumar:2016oop,Kumar:2016sxx} of a cosmological constant in the theory.

In this work, we will deepen our earlier studies 
\cite{Kleihaus:2009kr,Kumar:2017zms} of the domain of existence
of the {\it massless} theory with a conical potential
in terms of 2D phase diagrams.
Firstly we consider a larger range of the variables involved as compared to our earlier studies, and secondly we study additional 2D phase diagrams of the theory. We then extend these studies to the 3D phase diagrams, and we present the corresponding 3D plots of the various phase diagrams with a detailed discussion of the various regions of the phase diagrams with respect to the bifurcation points. 
In particular, we argue that there is an infinite sequence of such bifurcation points, and present an approximate formula for this sequence.
The theory is thus seen to contain rich physics in a particular domain of the phase diagrams. 

Further, we also study the distribution of the field variables of the theory with respect to the dimensionless radial coordinate at some representative points of the phase trajectories in the phase diagrams of the theory. For this purpose, we choose some particular phase trajectories taking them as examples to illustrate the properties of the corresponding compact boson stars. However, the procedure could be used to study the distribution of fields at any given representative point of any given trajectory (to be fixed by fixing the free dimensionless parameter of the theory designated by $a$ at a later stage in the text).

In our studies presented here, we construct the boson star solutions of this theory numerically. Our numerical procedure is based on the Newton-Raphson scheme with an adaptive stepsize Runge-Kutta method of order 4, and our numerical techniques have been calibrated by reproducing the previous work of Refs.~\cite{Kleihaus:2009kr,Kleihaus:2010ep,Kumar:2017zms} and \cite{Kumar:2014kna,Kumar:2015sia,Kumar:2016oop,Kumar:2016sxx}.

The paper is organized as follows. In the next section, we consider the mathematical formalism of the theory and derive the equations of motion. In Sec. \ref{sec:fe}, we rescale the constant parameters of the theory and the field variables to make them dimensionless. We further rescale the dimension-full radial coordinate $r$ to the dimensionless radial coordinate $\hat{r}$ 
and then express the equations of motion in terms of the dimensionless field variables as well as their derivatives expressed in terms of the dimensionless arguments. This then leads to a {\it single parameter theory}. In Sec. \ref{sec:bc}, we consider the boundary conditions. In Sec. \ref{sec:ns}, we lay down our scheme for obtaining the numerical solutions. In Sec. \ref{sec:2d} and \ref{sec:3d},  we study and discuss in detail the 2D and 3D phase diagrams of boson stars, respectively. In Sec. \ref{sec:mq}, we study the mass $~M$ and charge $~Q$ of these gravitating objects.  In Sec. \ref{sec:df}, we study the distribution of the fields with respect to the dimensionless radial coordinate of the theory for some representative points in the phase diagrams. The summary and conclusions are presented in Sec. \ref{sec:sc}. 

\section{Mathematical Formalism and Field Equations}\label{sec:mf}
The theory that we study is defined by the following action: 

\begin{eqnarray}
S&=&\int \left[ \frac{R}{16\pi G}   +\mathcal L_M \right] \sqrt{-g}\ d^4\,x\; ,\label{3:action}
\nonumber\\ \mathcal L_M& =&	- \frac{1}{4} F^{\mu\nu} F_{\mu\nu}
   -  \left( D_\mu \Phi \right)^* \left( D^\mu \Phi \right)
 - V(|\Phi|)\,,\nonumber\\
% F_{\mu\nu} &=& (\partial_\mu A_\nu - \partial_\nu A_\mu)\\, \nonumber\\\ \ 
 D_\mu \Phi &=& (\partial_\mu \Phi + i e A_\mu \Phi),\ \  \nonumber\\\ \ F_{\mu\nu} &=& (\partial_\mu A_\nu - \partial_\nu A_\mu) \,, \nonumber\\
 V(|\Phi|) &=& \lambda |\Phi| \,.
\end{eqnarray}
Here $R$ is the Ricci curvature scalar, and $G$ is Newton's gravitational constant,
$g = det(g_{\mu\nu})$, 
where $g_{\mu\nu}$ is the metric tensor, 
the asterisk denotes complex conjugation, and $\lambda$ is a constant parameter. The variational principle then leads to the equations of motion:
\begin{eqnarray}
G_{\mu\nu}&\equiv& R_{\mu\nu}-\frac{1}{2}g_{\mu\nu}R = 8\pi G T_{\mu\nu}\,,
\nonumber \\ 
 \partial_\mu \left ( \sqrt{-g} F^{\mu\nu} \right)&=&
   -i e \sqrt{-g}[\Phi^* (D^\nu \Phi)-\Phi (D^\nu \Phi)^* ]\,,
\nonumber 
\end{eqnarray}
\begin{eqnarray}
D_\mu\left(\sqrt{-g}  D^\mu \Phi \right)& =& \frac{\lambda }{2}\sqrt{-g}\,\frac{\Phi}{|\Phi|}\,,\nonumber \\
\left[D_\mu\left(\sqrt{-g}  D^\mu \Phi \right)\right]^*& =& \frac{\lambda }{2}\sqrt{-g}\,\frac{\Phi^*}{|\Phi|} ~, 
  \label{3:vfeqH}
 \end{eqnarray}
where the energy-momentum tensor $T_{\mu\nu}$ is given by
\begin{eqnarray}
T_{\mu\nu} &=& \biggl[ ( F_{\mu\alpha} F_{\nu\beta}\ g^{\alpha\beta} -\frac{1}{4} g_{\mu\nu} F_{\alpha\beta} F^{\alpha\beta})
\nonumber\\ & & + (D_\mu \Phi)^* (D_\nu \Phi)+ (D_\mu \Phi) (D_\nu \Phi)^*  
 \\ & &  -g_{\mu\nu} \left((D_\alpha \Phi)^* (D_\beta \Phi)    \right) g^{\alpha\beta}
 -  g_{\mu\nu} \lambda( |\Phi|) \biggr] . \ \ \  \nonumber \label{3:vtmunu}
\end{eqnarray}
We work with the static spherically symmetric metric 
in Schwarzschild-like coordinates
\begin{equation}
ds^2= \biggl[ -A^2 N dt^2 + N^{-1} dr^2 +r^2(d\theta^2 + \sin^2 \theta d\phi^2) \biggr] \,,
\end{equation}
where the arguments of the metric functions $A(r)$ and $N(r)$ have been suppressed. The components of Einstein tensor ($G_{\mu\nu}$) are then obtained as follows,
\begin{eqnarray}
G_t^t &=& \biggl[ \frac{-\left[r\left(1-N\right)\right]'}{r^2} \biggr] ,\nonumber \\  
G_r^r &=& \biggl[ \frac{2 r A' N -A\left[r\left(1-N\right)\right]'}{A\ r^2} \biggr] ,\nonumber \\
G_\theta^\theta &=& \biggl[ \frac{2r\left[rA'\ N\right]' + \left[A\ r^2 N'\right]'}{2 A\ r^2} \biggr]
\  \   = \  G_\varphi^\varphi \,,
\end{eqnarray}
where the prime denotes differentiation with respect to $~r$.

Assuming a vanishing magnetic field, we make the following Ans\"atze for the matter fields,
\begin{equation}
 \Phi(x^\mu)=\phi(r) e^{i\omega t}\ \ ,\ \ A_\mu(x^\mu) dx^\mu = A_t(r) dt .
\end{equation}
With these Ans\"atze the Einstein equations
\begin{eqnarray}
G_t^t = 8 \pi G\ T_t^t  \ ,\ \  \;&& G_r^r =  8 \pi G\  T_r^r  \ \,,\nonumber \\ 
G_\theta^\theta =  8 \pi G\ T_\theta^\theta \ ,\ \  &&G_\varphi^\varphi =  8 \pi G\ T_\varphi^\varphi
\end{eqnarray}
(with the arguments of the field variables being suppressed) reduce to:

\begin{strip}
\begin{eqnarray}
 \frac{-1}{r^2}\left[r\left(1-N\right)\right]' 
& = & \frac{-8\pi G}{2A^2 N e^2} \bigg[ N [(\omega +e A_t )']^2  +(\omega + e A_t )^2 (\sqrt{2} e \phi)^2 + A^2 N^2 (\sqrt{2} e \phi')^2 \left.+\frac{2\,e\,\lambda}{\sqrt 2}A^2N(\sqrt{2}\,e\,\phi)  \right]\,, \label{dtt} \\
\frac{2 r A' N -A\left[r\left(1-N\right)\right]'}{A r^2}
& = & \frac{8\pi G}{2A^2 N e^2}\bigg[ -N [(\omega +e A_t )']^2 +(\omega + e A_t )^2 (\sqrt{2} e \phi)^2+ A^2 N^2 (\sqrt{2} e \phi')^2  \left.- \frac{2\,e\,\lambda}{\sqrt 2}(\sqrt{2}A^2 N \,e\,\phi) \right]  \,,\label{drr} \\
\frac{2r\left[rA'N\right]' + \left[A r^2 N'\right]'}{2 A r^2} & = &\frac{8\pi G}{2A^2 N e^2} \bigg[ N [(\omega +e A_t )']^2 +(\omega + e A_t )^2 (\sqrt{2} e \phi)^2 -A^2 N^2 (\sqrt{2} e \phi')^2  \left.- A^2 N \frac{2\,e\,\lambda}{\sqrt 2}(\sqrt{2}\,e\,\phi)\right]   \label{dtheta} \,.
\end{eqnarray}
\end{strip}

\noindent  Here, the primes denote differentiation with respect to $~r$. Also, the equation $G_\varphi^\varphi =  8 \pi G\ T_\varphi^\varphi $ leads to an equation identical to Eq.~(\ref{dtheta}).

\section{Field Equations in Rescaled Variables}\label{sec:fe}

We now introduce the new constants,
\begin{equation}
 \beta=\frac{\lambda\,e}{\sqrt{2}}   \ \ \ ,\ \ \ \alpha^2 (~:=a) = \frac{4\pi G\,\beta^{2/3}}{e^2} , \label{parameters}
\end{equation}
where $~a$ is dimensionless, 
and redefine the functions $\phi(r)$ and $A_t(r)$ through the following relations:
\begin{equation}
 h(r)=\frac{(\sqrt{2} \;e\, \phi(r))}{\beta^{1/3}} \ \ \ ,\ \ \  b(r)=\frac{(\omega+e A_t(r))}{\beta^{1/3}} . \label{hb}
 \end{equation}
For the sake of completeness, we mention that the metric functions $A(r)$ and $N(r)$ are already dimensionless field variables.

We also introduce a dimensionless coordinate $\hat{r}$ 
defined by $\hat{r}:=\beta^{1/3}\,{r}$
(which in turn implies $\frac{d}{d{r}}=\beta^{1/3}\frac{d}{d\hat{r}}$).  
Eq.~(\ref{hb}) then gives:
\begin{equation}
 h(\hat{r})=\frac{(\sqrt{2} \;e\, \phi(\hat{r}))}{\beta^{1/3}} \ \ \ ,\ \ \  b(\hat{r})=\frac{(\omega+e A_t(\hat{r}))}{\beta^{1/3}} . \label{hb1} 
 \end{equation}
From now onwards, we will change the arguments of the field variables from the dimension-full radial coordinate $r$ to the dimensionless radial coordinate $\hat{r}$ and the primes will denote differentiation with respect to the dimensionless radial coordinate $\hat{r}$. % (instead of the dimension-full coordinate $r$). 
  
The matter field equations of motion involving the dimensionless field variables as well as their derivatives
%namely, the functions $~h(\hat{r})$ and $b(\hat{r})$ as well as the metric functions $A(\hat{r})$ and $~N(\hat{r})$ 
(with ${\rm sign }(h)$ denoting the usual signature function) then read:
\begin{eqnarray}
\left[A N \hat{r}^2 h'\right]' &  = & \frac{\hat{r}^2}{AN}\left(A^2N {\rm sign}(h) -b^2 h\right)\ ,  \label{3:vheq}\\
\left[\frac{\hat{r}^2 b'}{A}\right]' &  = & \frac{b h^2 \hat{r}^2}{AN}  . \  \label{3:vbeq}
\end{eqnarray}
We thus obtain the set of equations of motion 
to be solved numerically as:
\begin{eqnarray}
N' & = & \bigg[\frac{1-N}{\hat{r}} -\frac{\alpha^2 \hat{r}}{A^2 N}\left(A^2 N^2 h'^2 + N b'^2 \right.\nonumber\\ & &\hspace{1in}\left.+2 A^2 N h+ b^2 h^2\right)\bigg] \ , \label{eq_N}\\
A' & = & \bigg[\frac{\alpha^2 \hat{r}}{A N^2}\left(A^2 N^2 h'^2 + b^2 h^2\right)\bigg] ,\\ \label{eq_A}
h'' & = &\bigg[ \frac{\alpha^2}{A^2N} \hat{r} h' \left(2A^2  h +b'^2\right)-\frac{h'\left(N+1\right)}{\hat{r} N}\nonumber\\&&\qquad+\frac{A^2N {\rm sign}(h)-b^2 h}{A^2 N^2} \bigg], \label{eq_H}\\
b'' & = & \bigg[\frac{\alpha^2}{A^2 N^2} \hat{r}b'\left(A^2 N^2 h'^2 + b^2 h^2\right)-\frac{2 b'}{\hat{r}} + \frac{b h^2}{N}\bigg] . \label{eq_b}
\end{eqnarray}

In order to solve the above equations of motion numerically, we further introduce a new coordinate $x$ defined through the following relation:

\begin{equation}
\hat{r} = x ~\hat{r}_o, ~~ 0\le x \le 1,
\end{equation}
implying that 
%$~\hat{r} = 0 $ at $ ~x = 0$ and $\hat{r} = \hat{r}_o$ at $~x = 1$.
the center of the boson star i.e. $\hat{r}=0$ corresponds to $x = 0$ and the outer
boundary of the boson star, i.e.~$\hat{r}=\hat{r}_o$ corresponds to $x = 1$.

\section{Boundary Conditions}\label{sec:bc}

We now need to determine the boundary conditions for the functions describing the metric and the fields.
For the metric function $A(\hat{r})$ we choose
\begin{equation}
A(\hat{r}_o)=1 \label{Aro} \\,
%\label{abc}
\end{equation}
where $\hat{r}_o$ denotes the outer radius of the star. 
Thus $A$ assumes its asympototic value already at the star's boundary.

In order to obtain globally regular ball-like boson star solutions, we impose
\begin{eqnarray}
& N(0)=1 \ ,\ \   b'(0)=0 \ ,\nonumber\\ &h'(0)=0\ ,\ \ h(\hat{r}_o)=0\ , \ \  h'(\hat{r}_o)=0 . \label{bcstar}
\end{eqnarray}
Since the star has a sharp boundary, with the electro-vac equations holding
in the exterior region $\hat{r}>\hat{r}_o$, we match
the Reissner-Nordstr\"om solution at  $\hat{r}=\hat{r}_o$.

The U(1) invariance of the theory leads to
a conserved Noether current,
\begin{eqnarray}
j^\mu=-i \,e\,\left[ \Phi(D^\mu \Phi)^*-\Phi^* (D^\mu \Phi) \right]\ ,\   \
j^{\mu}_{\ ;\mu} = 0\,,
\end{eqnarray}
whose time component corresponds to the charge density.
The global charge $Q$ of the boson star is then given by
\begin{equation}
Q=-\frac{1}{4\pi}\int_{0} ^{\hat{r}_o} j^t \sqrt{-g} \,dr\,d\theta\,d\phi  \,,\  
j^t=-\frac{h^2(\hat{r}) b(\hat{r})}{A^2(\hat{r}) N(\hat{r})} .\label{charge} \nonumber
\end{equation}
We like to clarify here that the frequency of the scalar field $\omega$ shows up explicity in the Eqs. (\ref{dtt})-(\ref{dtheta}) however, it does not show up explicitley in Eqs.(\ref{3:vheq})-(\ref{eq_b}), merely because we introduced new dimensionaless field variables (with dimensionless arguments) through Eqs. (\ref{hb})-(\ref{hb1}). and the definition of $b(\hat{r})$ contains $\omega$ explicitly.
The charge density given by Eq. (\ref{charge}) is seen to be proportional to $b(\hat{r})$ which in turn has its definition (in terms of $\omega$) given by Eqs. (\ref{hb})-(\ref{hb1}). One could thus visualize the dependence of the charge or of the charge density in terms of $\omega$ if one likes.

The mass can be read off the metric, as usual, making use of the fact,
that the metric outside the compact star corresponds to a 
Reissner-Nordstr\"om metric.
In the units employed, we then obtain for the mass $M$ of the boson star solutions
 \begin{equation}
M= \biggl(1-N(\hat{r}_o)+\frac{\alpha^{2} Q^{2}}{\hat{r}_o^2}\biggr)\frac{\hat{r}_o}{2} .
\end{equation}

An extremal Reissner-Nordstr\"om metric would satisfy a proportionality between
the mass $~M$~ and the charge $~Q~$ given by $~M = \alpha Q = \sqrt{a} Q~$, 
where $~ \alpha (:= \sqrt{a}) ~$ enters here because of the units employed
(whereas in the usual geometric units the extremal solution satisfies $~M=Q~$).
We can now consider three different cases for the exterior solution
($r>r_o$) outside the bosonic matter distribution:
(i) The case $~M/Q < \sqrt{a}~$ corresponds to 
a horizonless (naked) Reissner-Nordstr\"om solution.
(We note, that naked solutions would also arise, 
when the mass and charge of known elementary particles 
would be inserted on the left hand side of this relation.)
(ii) The case $~M/Q = \sqrt{a}~$ corresponds to 
an extremal Reissner-Nordstr\"om solution. 
(We note, that the sets of boson star solutions end,
when such an extremal Reissner-Nordstr\"om solution 
is reached and $b(0)=0$.)
(iii) The case $~M/Q > \sqrt{a}~$ corresponds to
a solution where the radius of the boson star is greater 
than its putative event horizon radius $~{\hat{r}_H}~$ 
(implying that the boson star is well outside the range 
of becoming a black hole. 

This discussion is reflected in Figs.~\ref {fig8},  
which show the variation of the various fields 
with respect to $~\hat{r} ~$. Here it is evident that the metric function 
$~N(\hat{r})~$ of the (selected) solutions 
always has a non-zero positive value,
(and also does not become zero at the outer radius of the boson star) 
implying thereby that the boson stars either have a radius 
$~\hat{r}  >  {\hat{r}_H}~$, 
or that they are smoothly matched
to a naked Reissner-Nordstr\"om solution for $~M/Q < \sqrt{a}~$.
The case $~M/Q = \sqrt{a}~$ is not exhibited in Figs.~\ref {fig8}.
It would correspond to a solution where a throat develops
and $b(0)=0$.

In the present work, we study (i) the variation of $M $ with $\hat r_0 $,
(ii) the variation of $Q$ with $\hat r_0 $,
(iii) the variation of $M$ with $Q$,
and (iv) the variation of $M/Q$ with $Q$, 
where the dimensionaless variable $~a~$ varies in the range 
$a = 0.050 $ to $a = 0.250 $  
(to be seen in detail later in the discussion of the results). 
 
\begin{figure*}
\begin{center} 
	\mbox{\subfigure[][]{\includegraphics[scale=0.68]{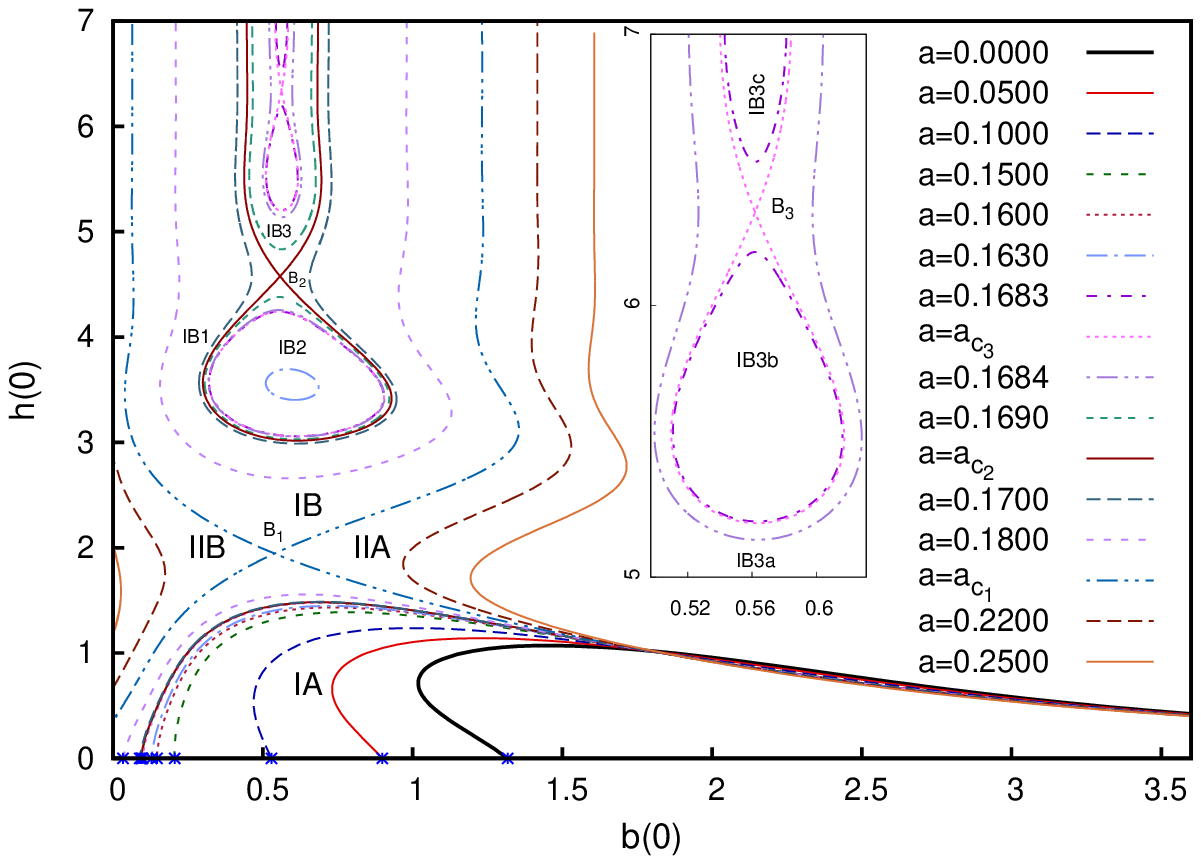}\label{f1a}}
		  \subfigure[][]{\includegraphics[scale=0.68]{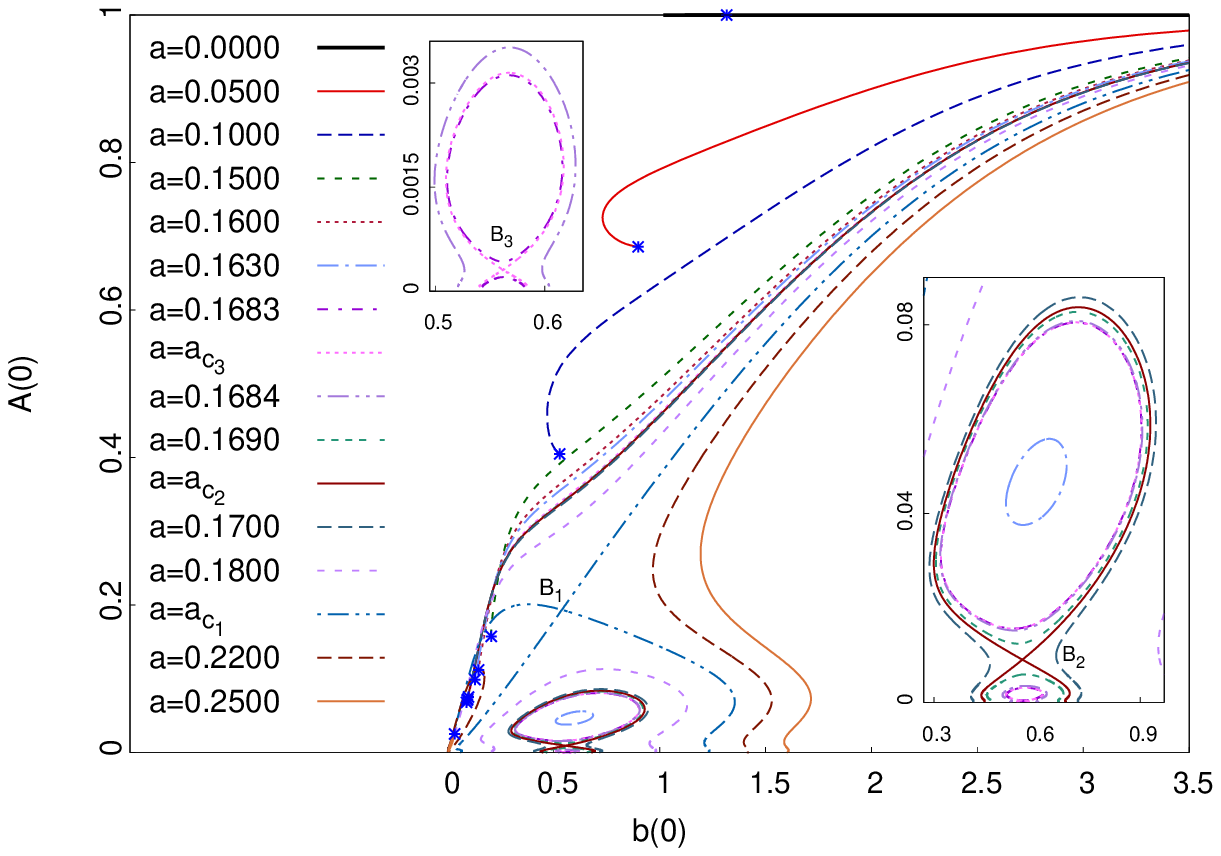}\label{f1b}}}
    \mbox{\subfigure[][]{\includegraphics[scale=0.68]{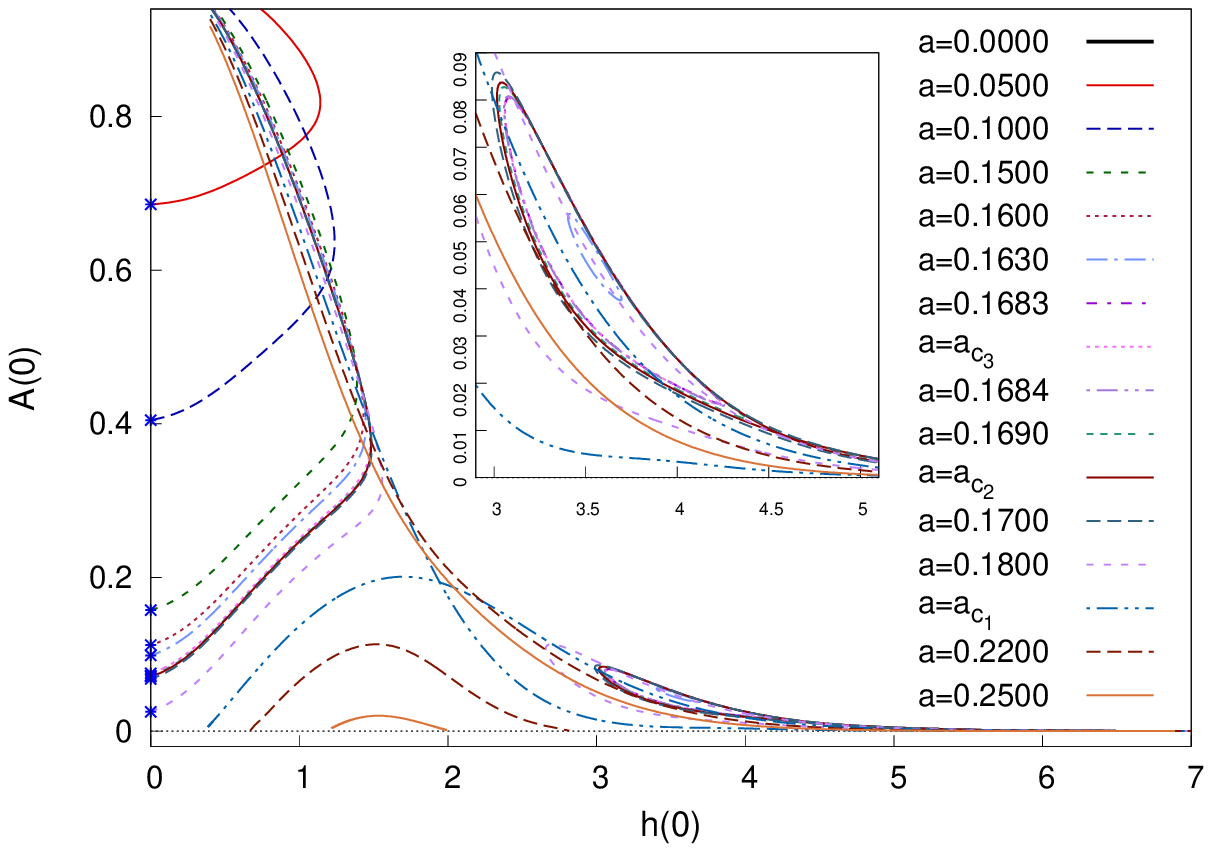}\label{f1c}}
    	  \subfigure[][]{\includegraphics[scale=0.68]{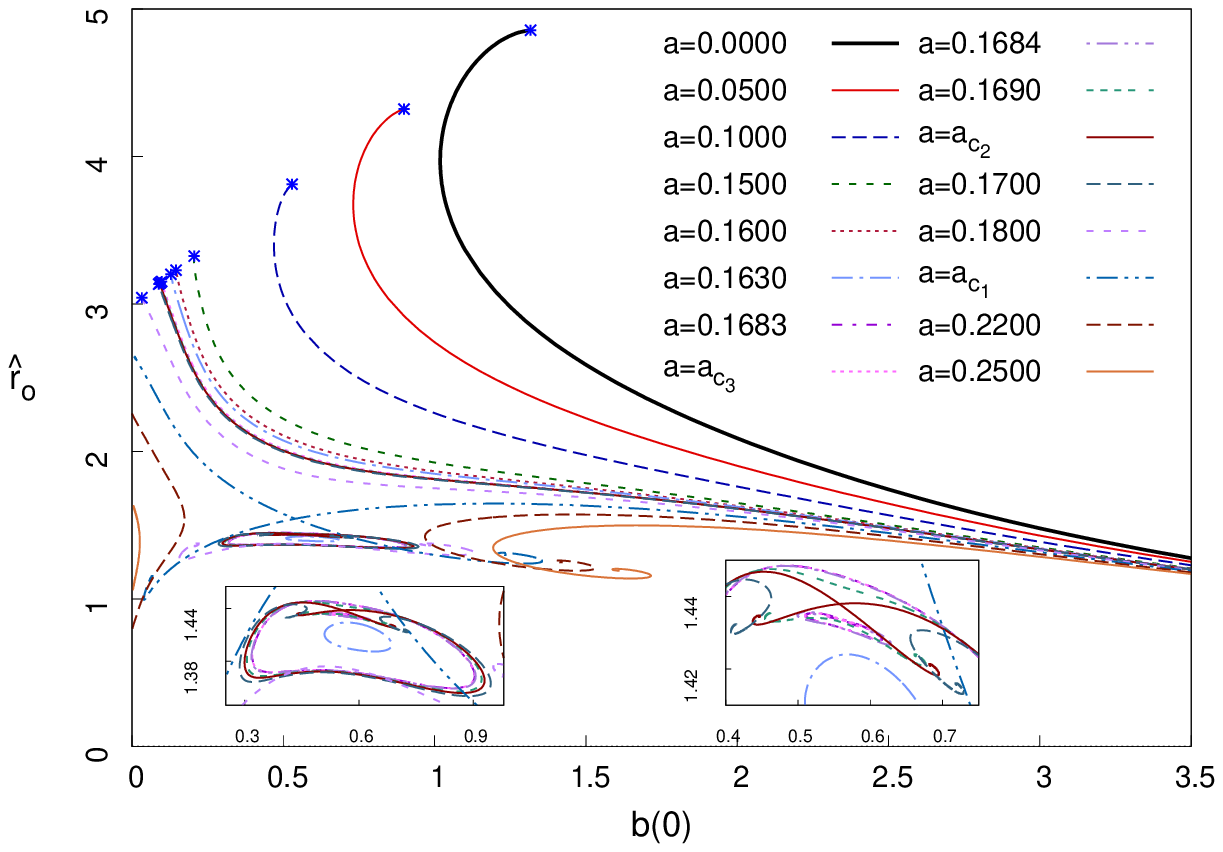}\label{f1d}}}
 	\caption{Figs.~(a) to (c) depict the 2D phase diagrams, showing, respectively, $h(0)$ versus $b(0)$, $A(0)$ versus $b(0)$ and $A(0)$ versus $h(0)$ (where $h(0)$, $b(0)$ and $A(0)$  denote the values of these fields at the centre of the star). Fig.~(d) depicts $\hat{r}_o$ versus $b(0)$. The figures show different sets of $a$ in the range $a=0.0$ to $a =0.2500$. The insets in Figs.~(a)--(d) magnify particular interesting areas of the figures. The asterisks represent the transition points from the boson stars to boson shells.\label{fig1}}
\end{center}
\end{figure*}

\begin{figure}
\begin{center}
        \mbox{\includegraphics[scale=0.68]{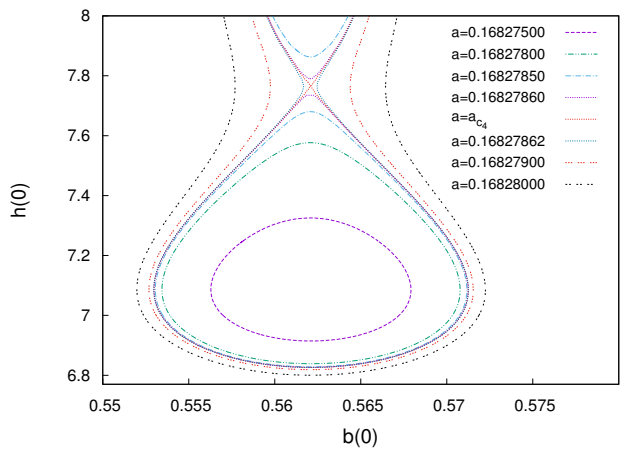}\label{f1a1}}
\caption{Zoom and extenstion of the 2D phase diagram Fig.~\ref{f1a}, showing $h(0)$ versus $b(0)$ into the region of the 4th bifurcation point $a_{c_4}$.
}
\end{center}
\end{figure}

\begin{figure*}
\begin{center}
	\mbox{\subfigure[][]{\includegraphics[scale=0.68]{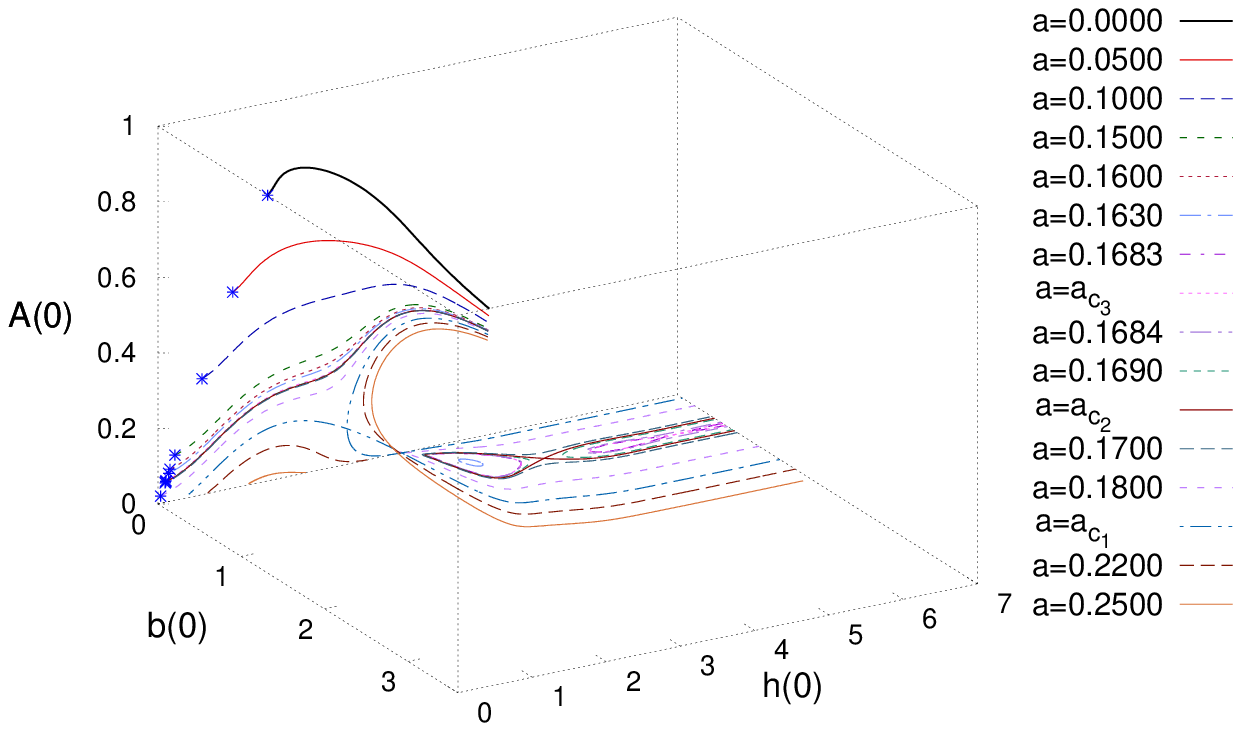}\label{f2a}}
		  \subfigure[][]{\includegraphics[scale=0.68]{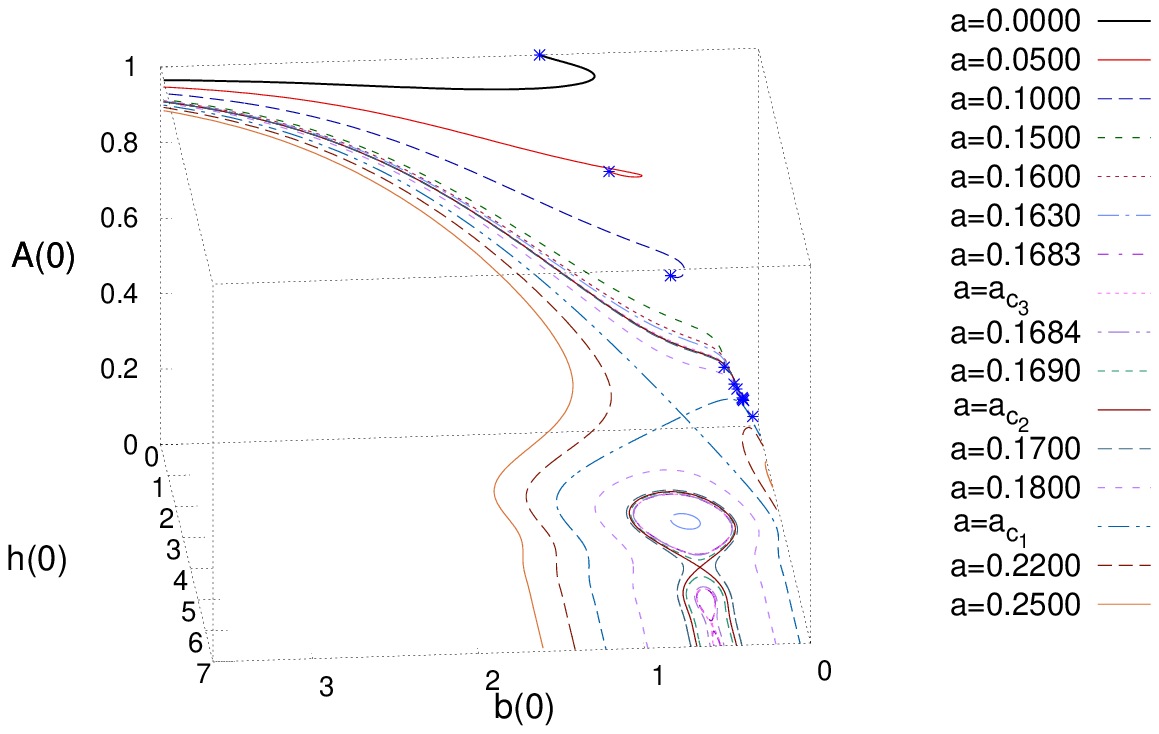}\label{f2b}}}
	\mbox{\subfigure[][]{\includegraphics[scale=0.68]{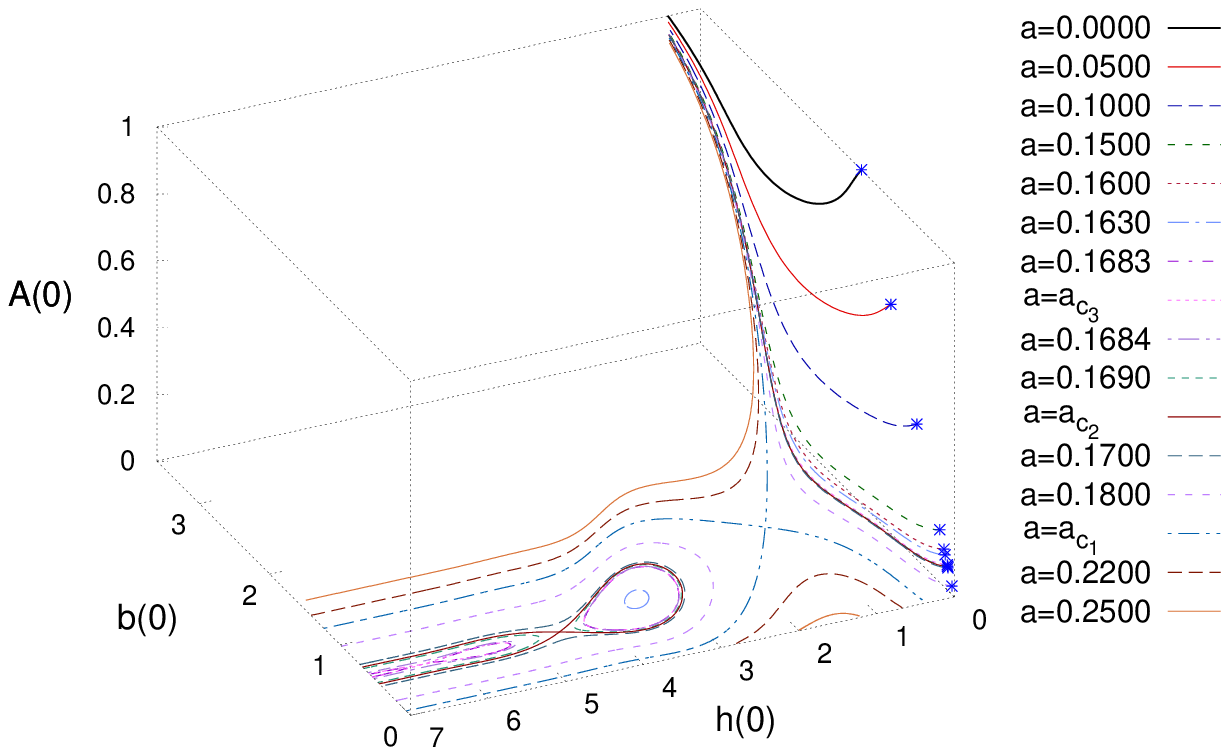}\label{f2c}}
		  \subfigure[][]{\includegraphics[scale=0.68]{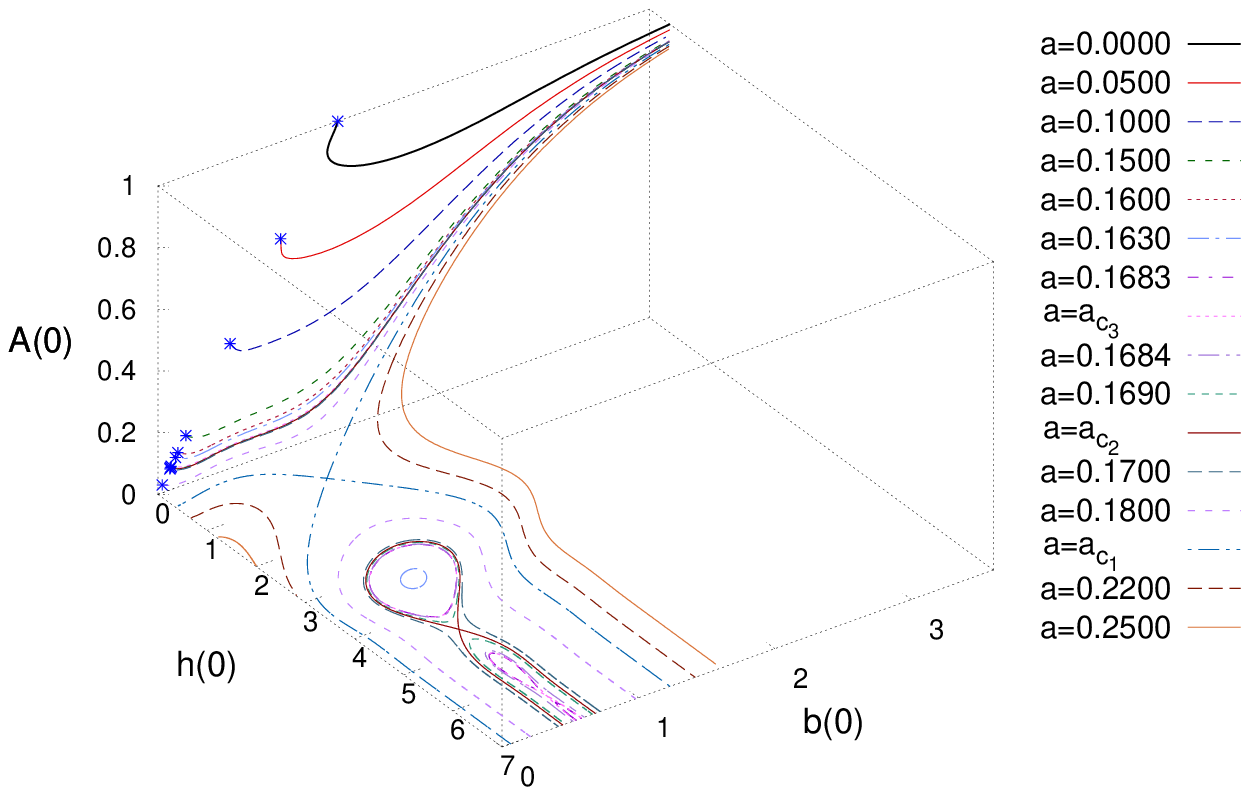}\label{f2d}}}
	\mbox{\subfigure[][]{\includegraphics[scale=0.68]{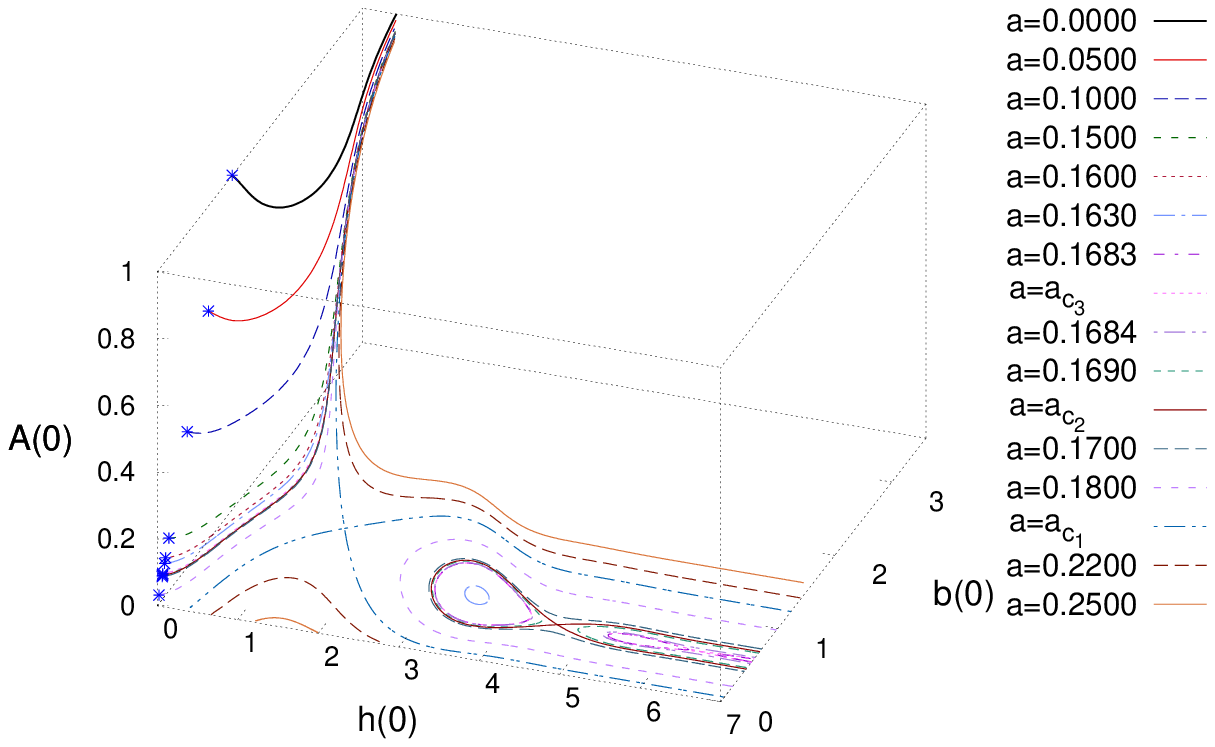}\label{f2e}}
		  \subfigure[][]{\includegraphics[scale=0.68]{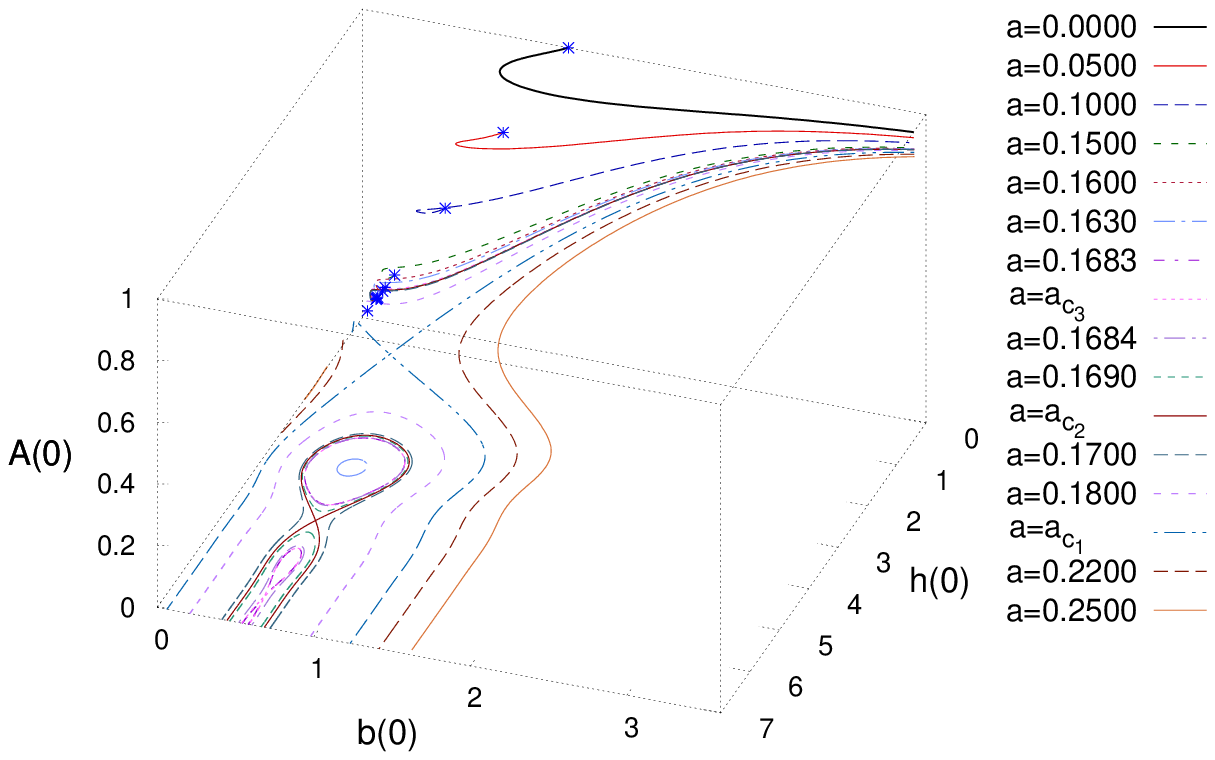}\label{f2f}}}
	\caption{Figs.~(a) to (f) show the 3D phase diagram (for different viewing angles) of $h(0)$ and 
$b(0)$ versus $A(0)$ for a set of values of $a$ in the range $a=0.0$ to $a =0.2500$.  The viewing angles are denoted by $(\theta, \phi)$, where $\theta$ is the angle of rotation about the $ox$-axis in the anticlockwise direction, which can take values between 0 to $\pi$, and $\phi$ is the angle of rotation about the $oz^\prime$-axis also in the anticlockwise direction, which can take values between 0 to $2\pi$. Figs.~(a)-(f) correspond, respectively, to the values $(\theta,\phi)\equiv(60,60),\ (60,175),\ (50,245),\ (135,35),\ (130,70) \mbox{ and } (135,340)$.\label{fig2}}
\end{center}
\end{figure*}

\begin{figure*}
\begin{center}
	\mbox{\subfigure[][]{\includegraphics[scale=0.68]{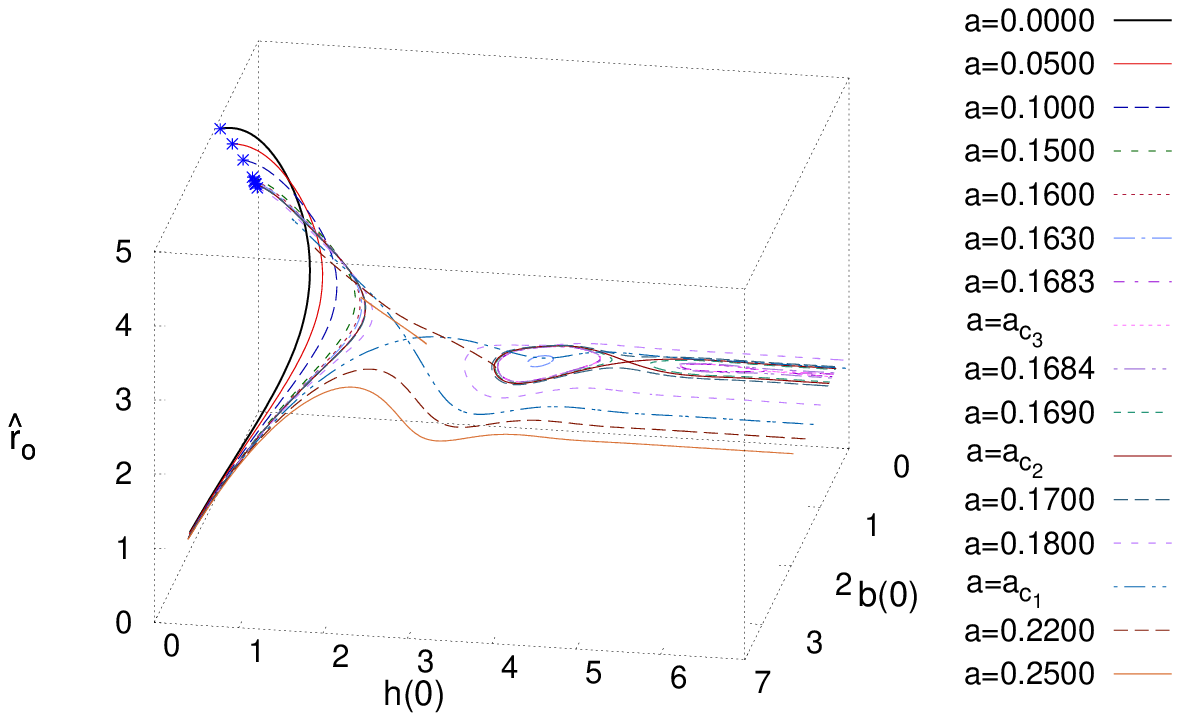}\label{f3a}}
		  \subfigure[][]{\includegraphics[scale=0.68]{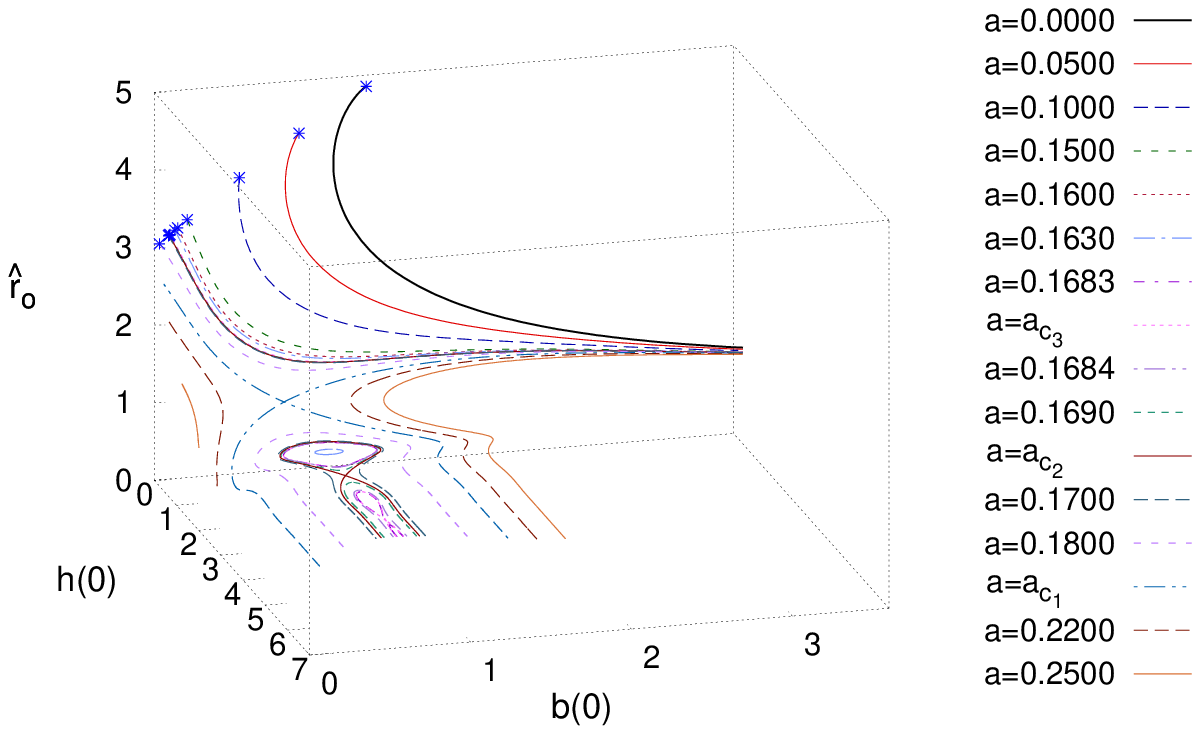}\label{f3b}}}
	\mbox{\subfigure[][]{\includegraphics[scale=0.68]{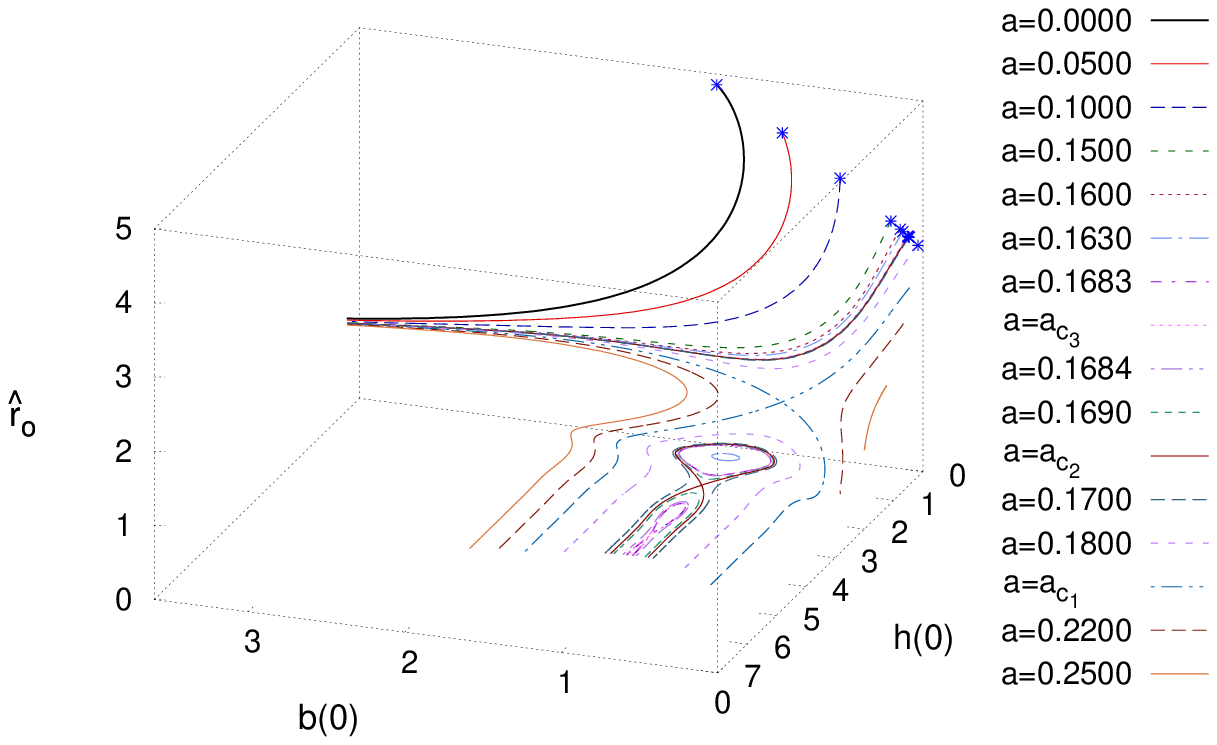}\label{f3c}}
		  \subfigure[][]{\includegraphics[scale=0.68]{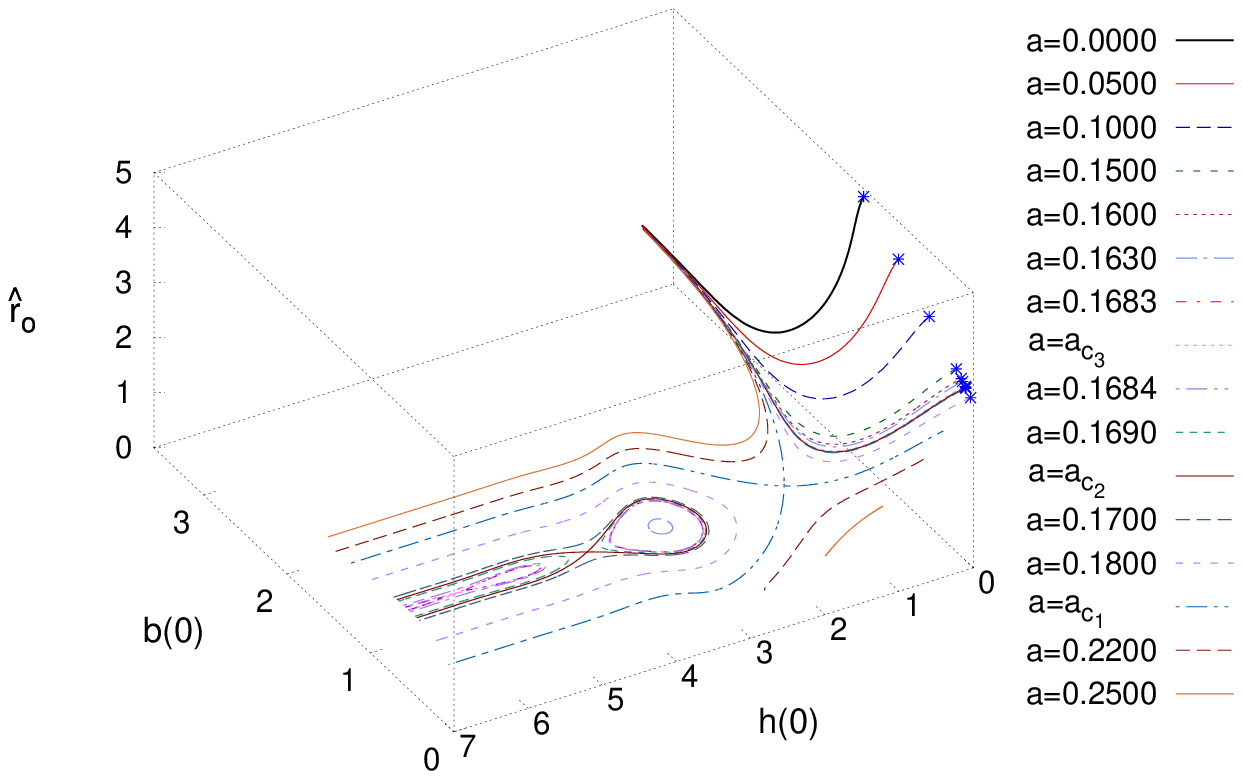}\label{f3d}}}
	\mbox{\subfigure[][]{\includegraphics[scale=0.68]{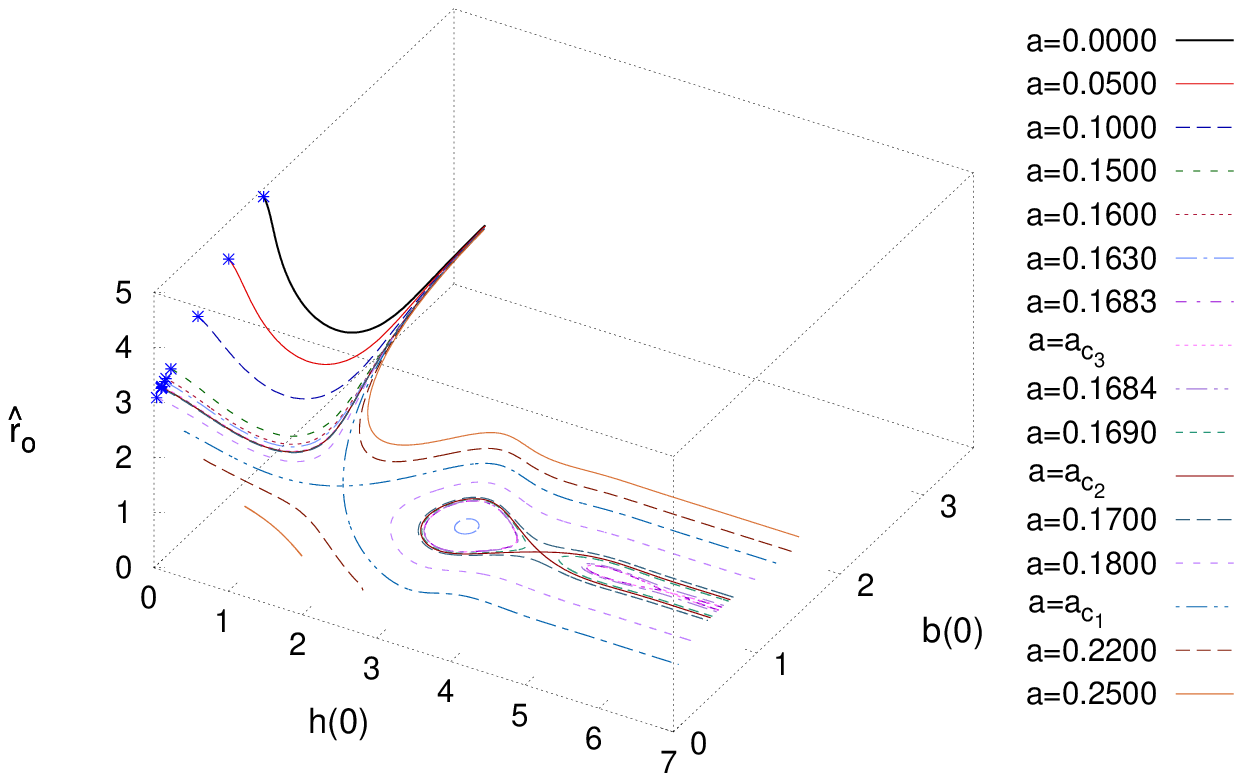}\label{f3e}}
		  \subfigure[][]{\includegraphics[scale=0.68]{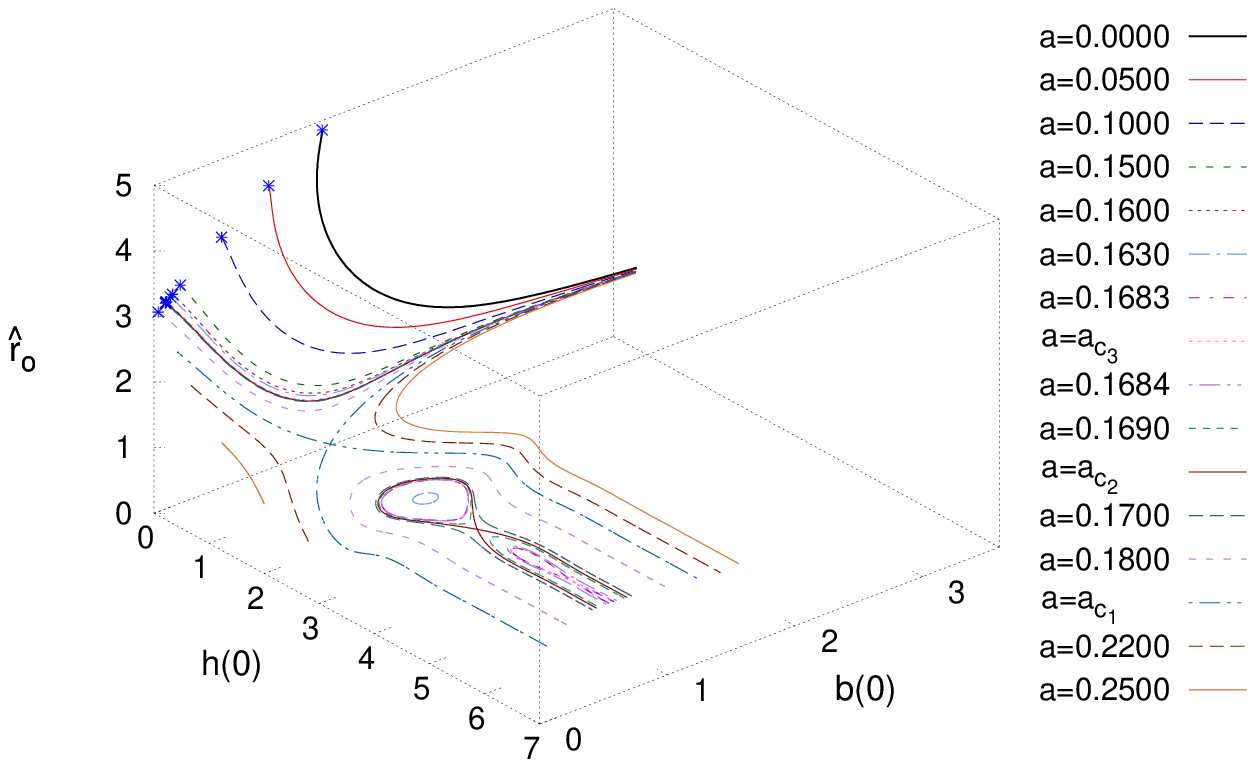}\label{f3f}}}
	\caption{
Figs.~(a) to (f) show the 3D phase diagram (for different viewing angles) of $h(0)$ and 
$b(0)$ versus $\hat r_o$ for a set of values of $a$ in the range $a=0.0$ to $a =0.2500$.  The viewing angles are denoted by $(\theta, \phi)$, where $\theta$ is the angle of rotation about the $ox$-axis in the anticlockwise direction, which can take values between 0 to $\pi$, and $\phi$ is the angle of rotation about the $oz^\prime$-axis also in the anticlockwise direction, which can take values between 0 to $2\pi$. Figs.~(a)-(f) correspond, respectively, to the values $(\theta,\phi)\equiv(60,100),\ (115,15),\ (60,200),\ (40,240),\ (140,60) \mbox{ and } (130,40)$.\label{fig3}}
\end{center}
\end{figure*}
\begin{figure}
\begin{center}
	\mbox{\subfigure[][]{\includegraphics[scale=0.66]{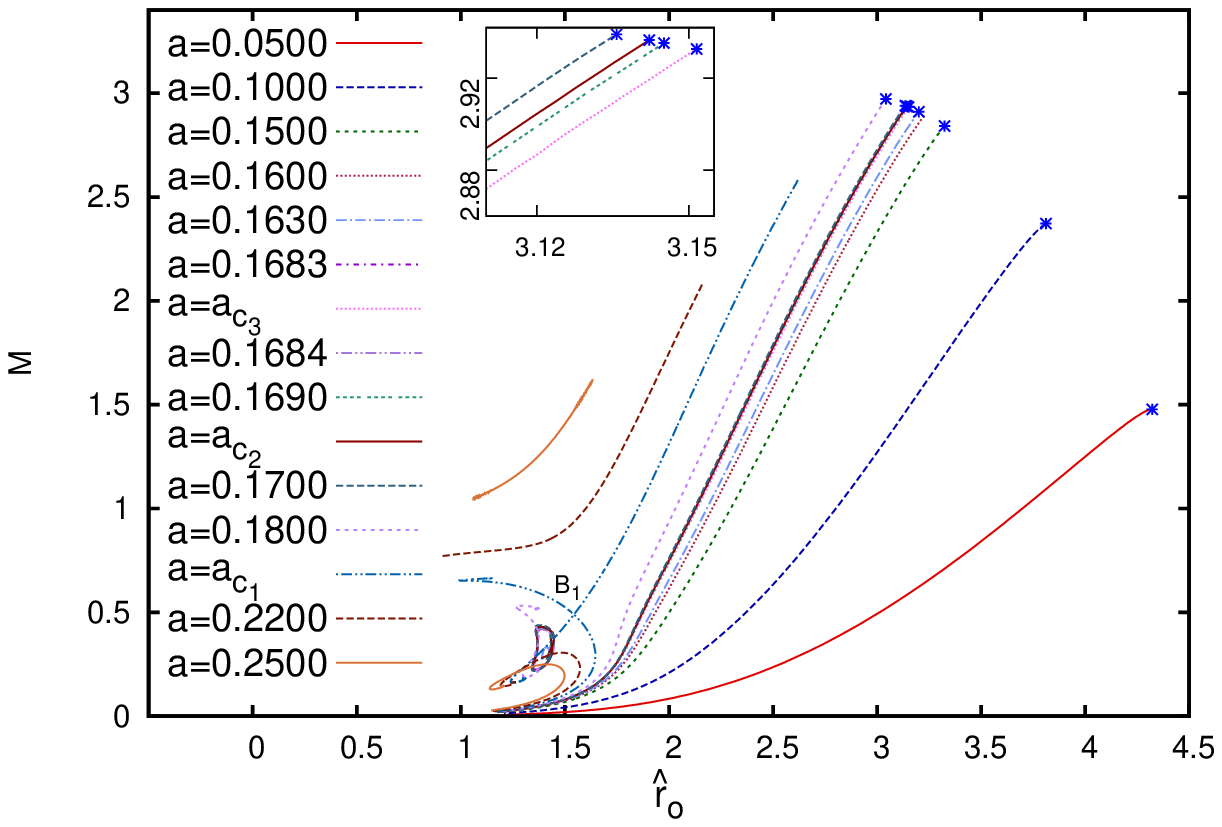}\label{f4a}}}
	\mbox{\subfigure[][]{\includegraphics[scale=0.66]{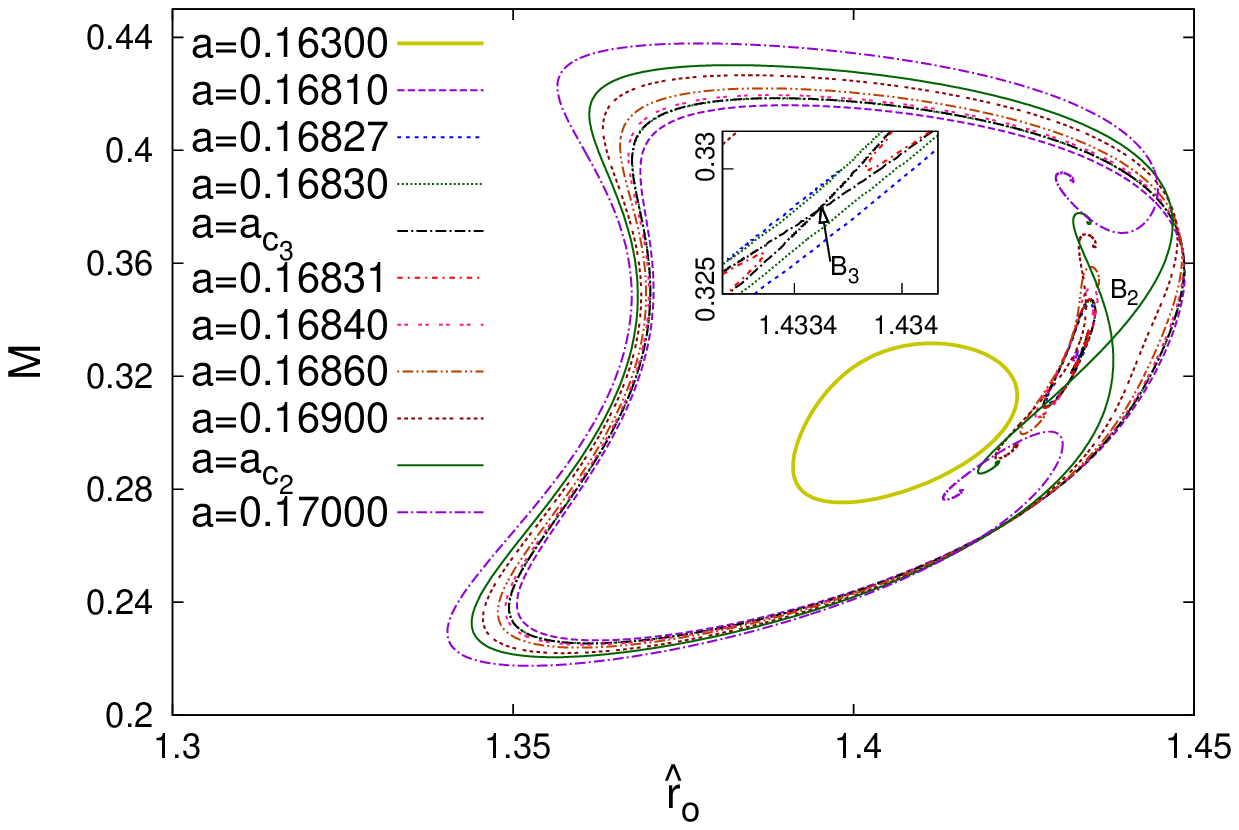}\label{f4b}}}
 	\caption{Fig.~(a) depicts the mass $M$ versus the radius of the star $\hat{r}_{o} $ for a set of values of $a$ in the range $a=0.0$ to $a =0.2500$. The asterisks represent the transition points from the boson stars to boson shells, and the insets magnify parts of the diagram. Fig.~(b) zooms into the region of the bifurcations, with the inset giving a magnified view of the bifurcation B3.\label{fig4}}
\end{center}
\end{figure}
\begin{figure}
\begin{center}
	\mbox{\subfigure[][]{\includegraphics[scale=0.66]{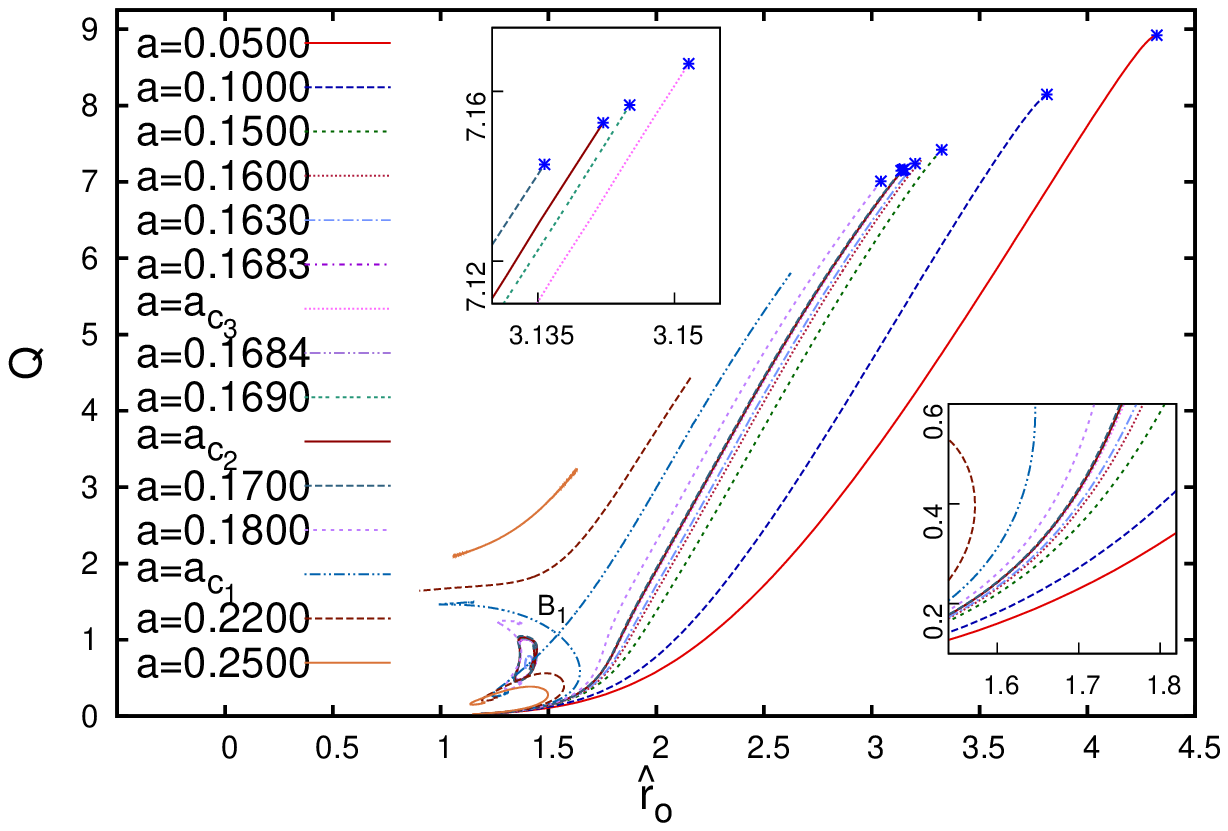}\label{f5a}}}
    \mbox{\subfigure[][]{\includegraphics[scale=0.66]{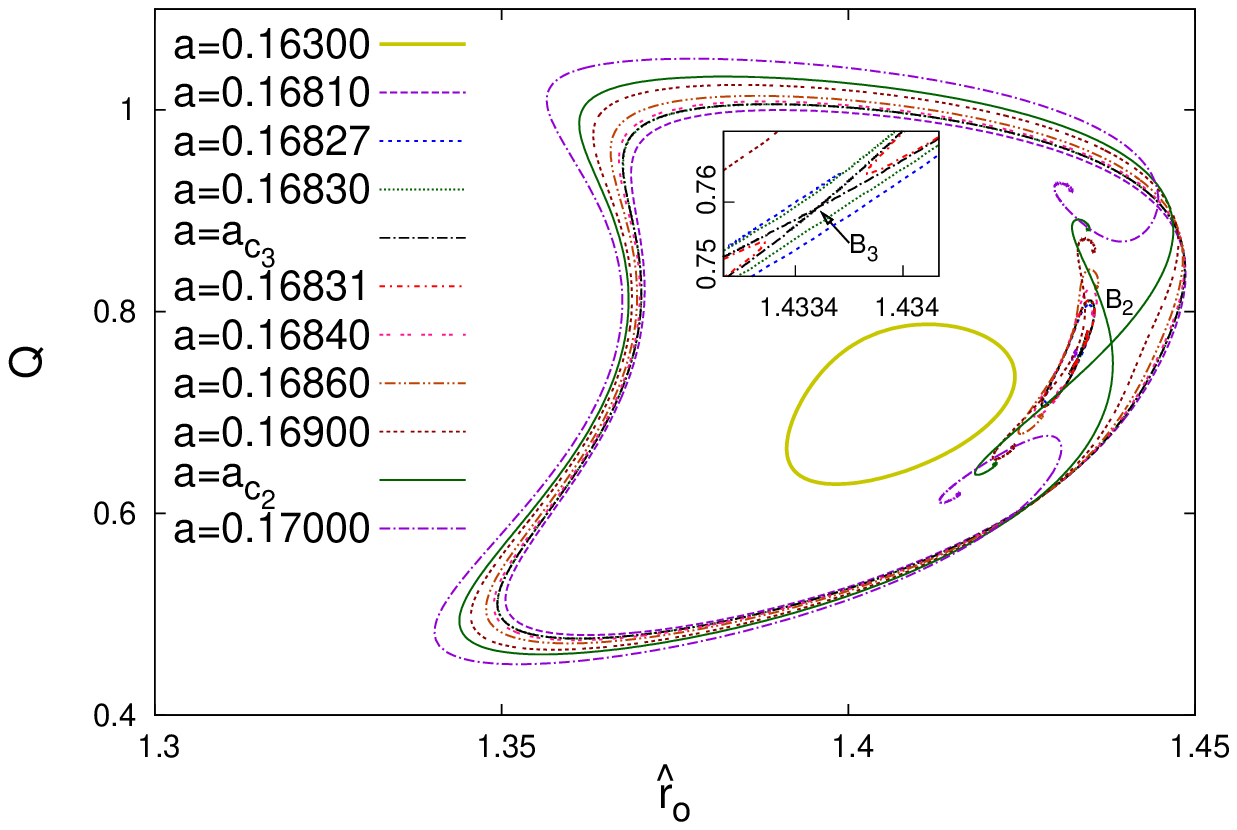}\label{f5b}}}
 	\caption{Fig.~(a) depicts the Charge $Q$ versus the radius of the star $\hat{r}_{o} $ for the same sequence of values of the parameter $a$. The asterisks represent the transition points from the boson stars to boson shells, and the insets magnify parts of the diagram. Fig.~(b) zooms into the region of the bifurcations, with the inset giving a magnified view of the bifurcation B3.\label{fig5}}
\end{center}
\end{figure}
\begin{figure}
\begin{center}
	\mbox{\subfigure[][]{\includegraphics[scale=0.66]{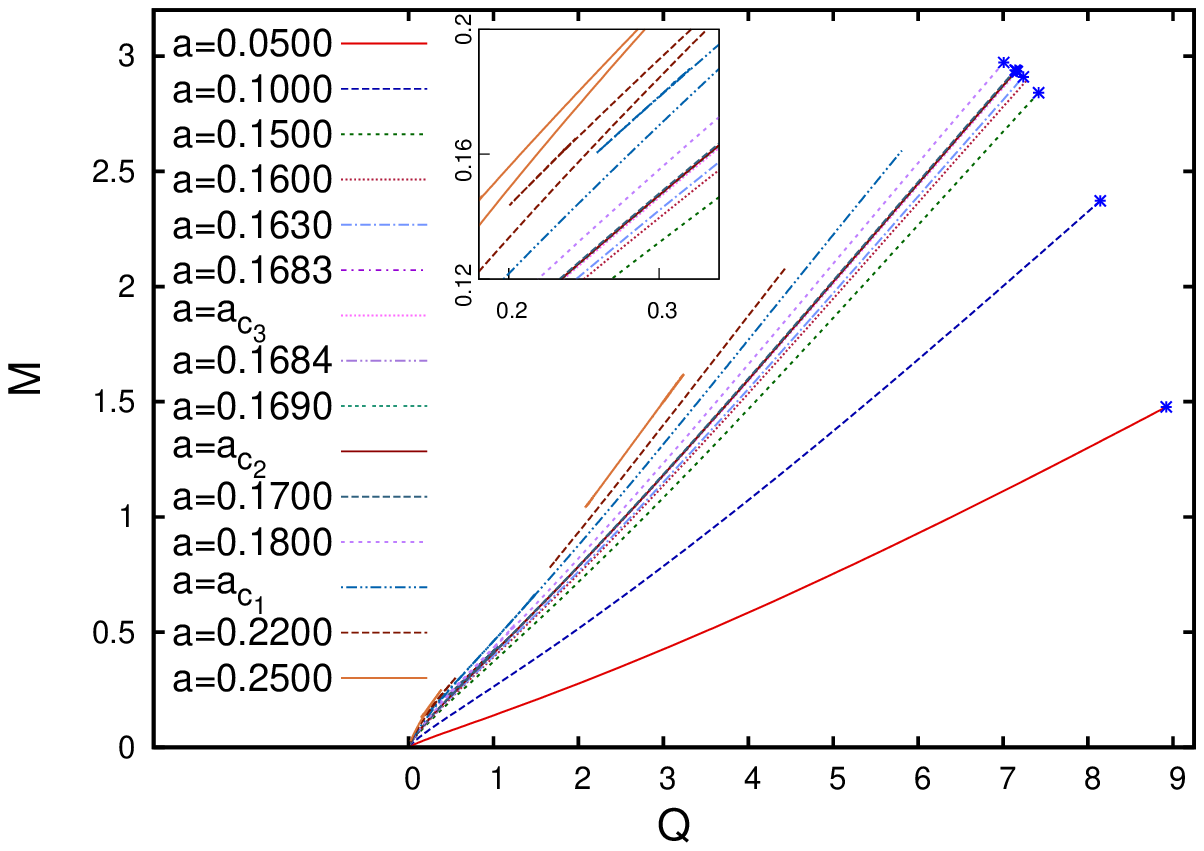}\label{f6a}}}
	\mbox{\subfigure[][]{\includegraphics[scale=0.66]{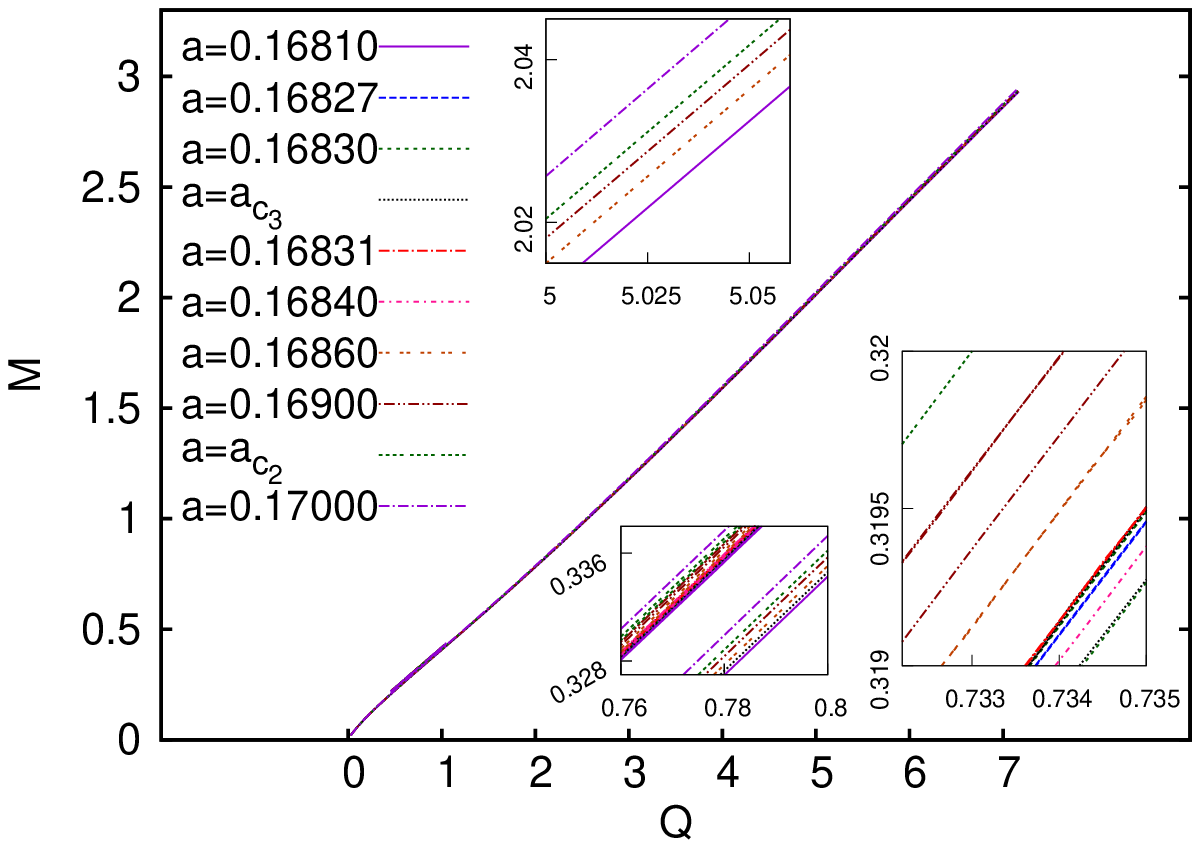}\label{f6b}}}
    \mbox{\subfigure[][]{\includegraphics[scale=0.66]{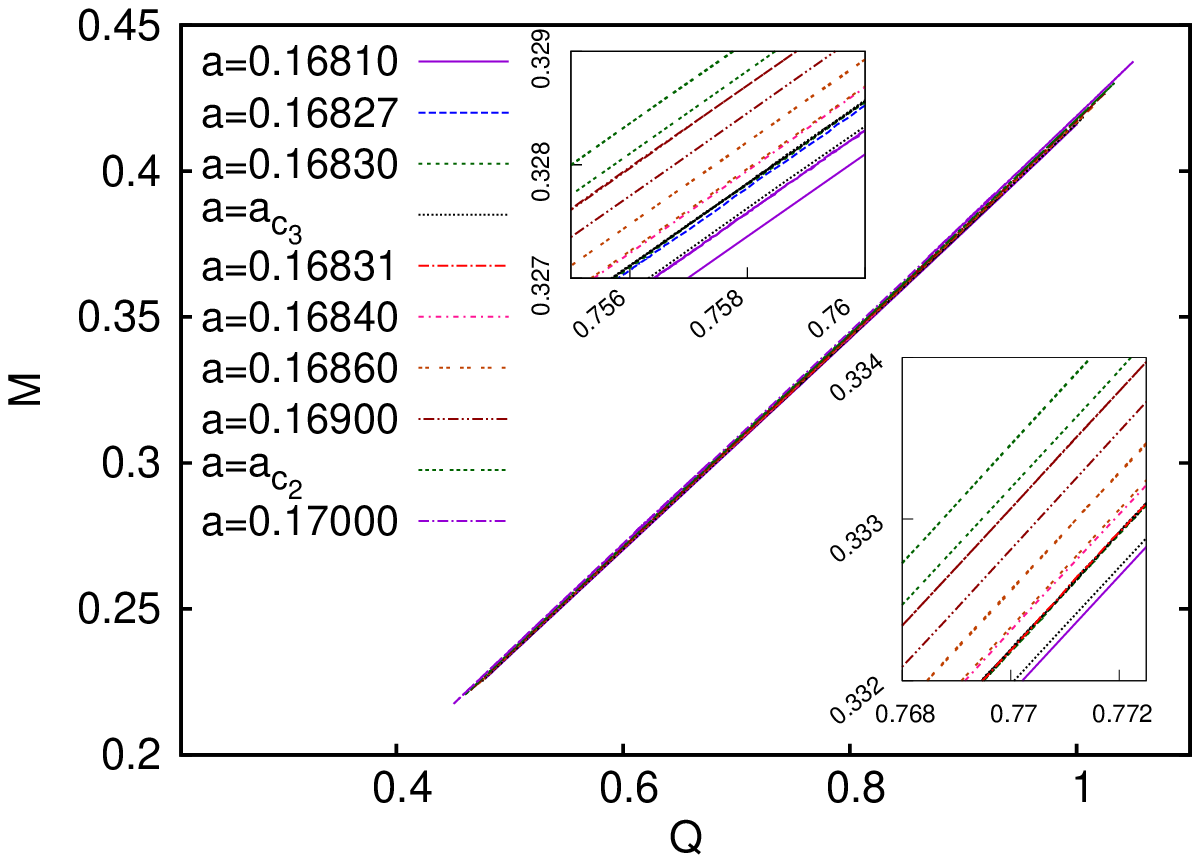}\label{f6c}}}
 	\caption{Figs.~(a) to (c) depict the mass $M$ versus the charge $Q$ for the same set of solutions. The asterisks represent the transition points from the boson stars to boson shells, and the inset magnifies a part of the diagram. \label{fig6}}
\end{center}
\end{figure}
\begin{figure}
\begin{center}
	\mbox{\subfigure[][]{\includegraphics[scale=0.66]{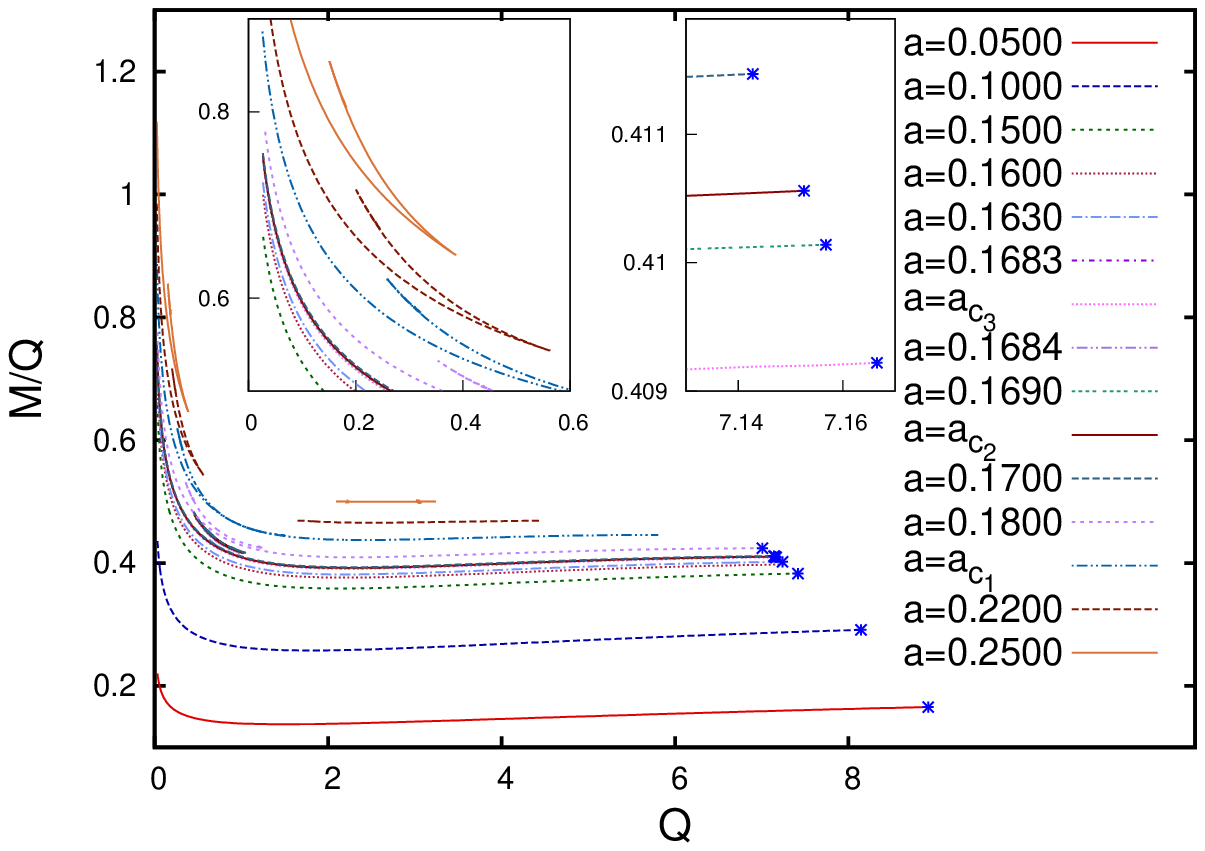}\label{f7a}}}
	\mbox{\subfigure[][]{\includegraphics[scale=0.66]{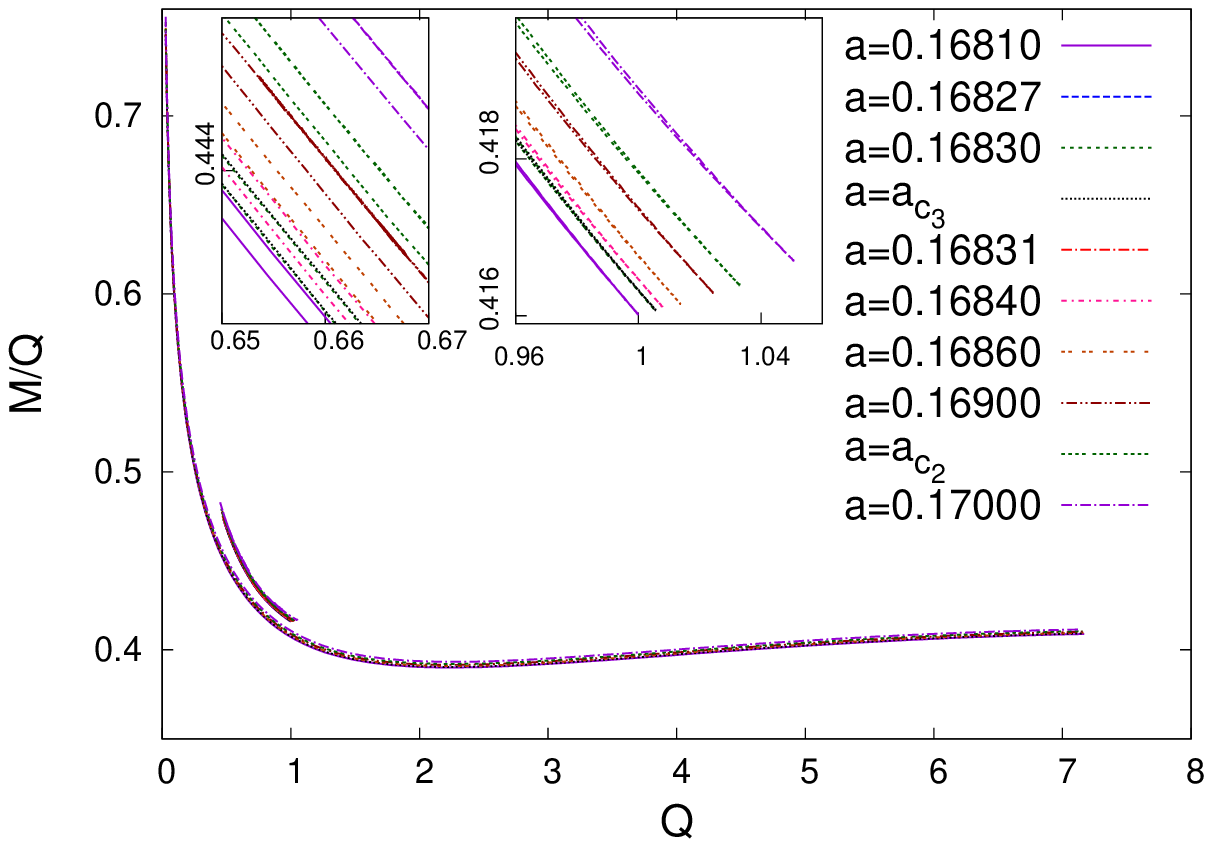}\label{f7b}}}
    \mbox{\subfigure[][]{\includegraphics[scale=0.66]{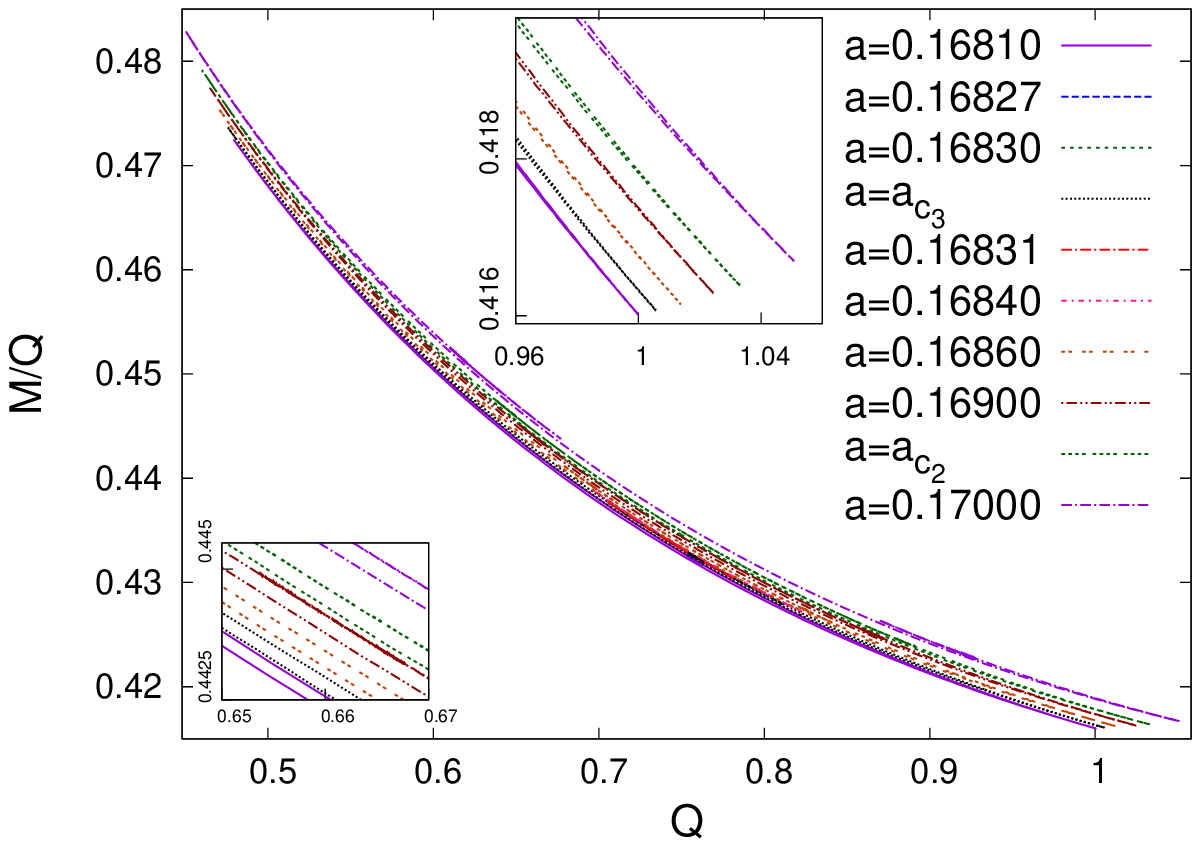}\label{f7c}}}
 	\caption{Figs.~(a) to (c) depict the mass per unit charge $M/Q$ versus the charge $Q$ for the same set of solutions. The asterisks represent the transition points from the boson stars to boson shells, and the inset magnifies a part of the diagram. \label{fig7}}
\end{center}
\end{figure}
\begin{figure*}
\begin{center}
	\mbox{\subfigure[][]{\includegraphics[scale=0.68]{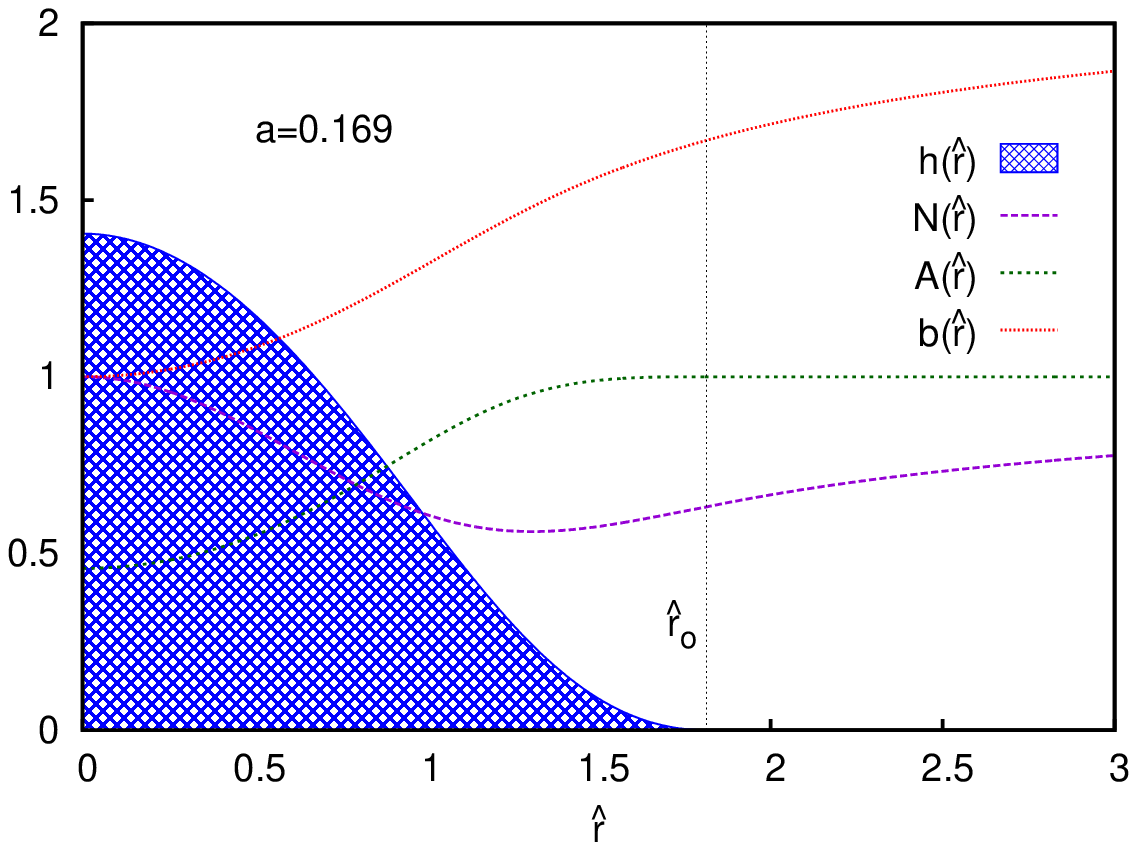}\label{f8a}}
		  \subfigure[][]{\includegraphics[scale=0.68]{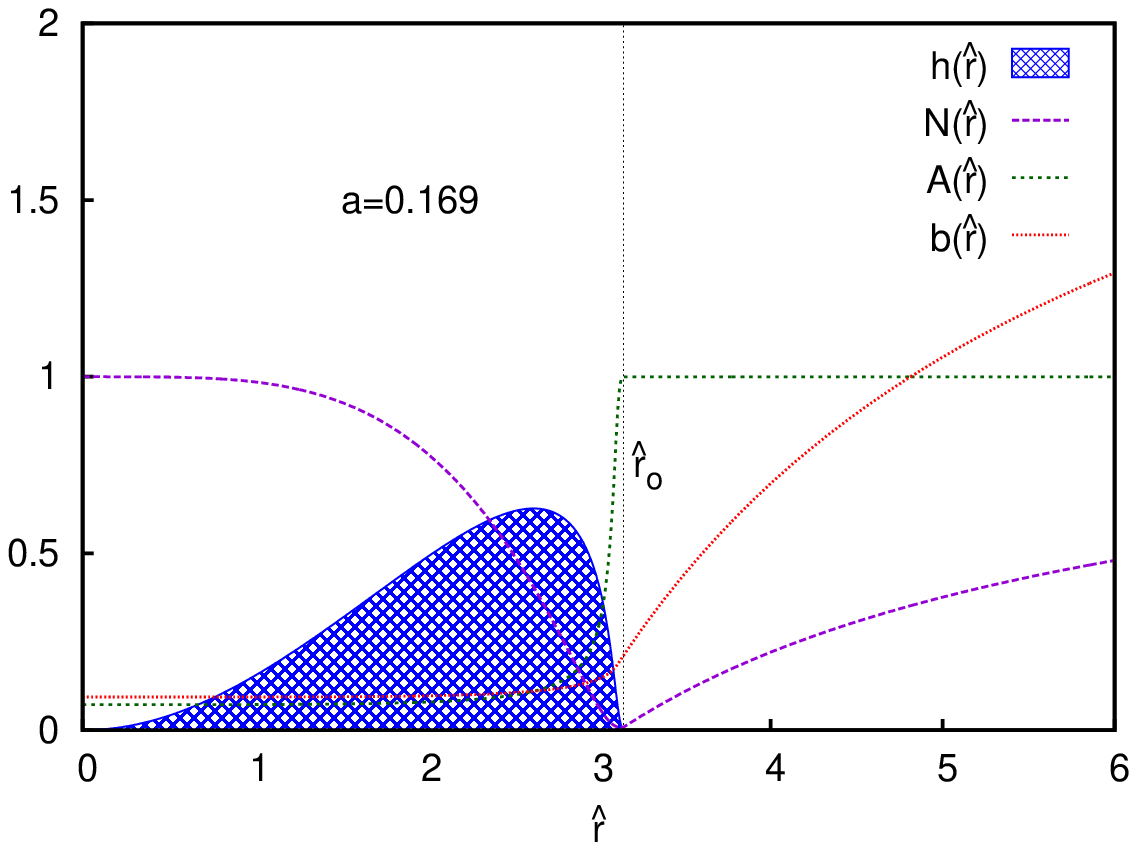}\label{f8b}}}
	\mbox{\subfigure[][]{\includegraphics[scale=0.68]{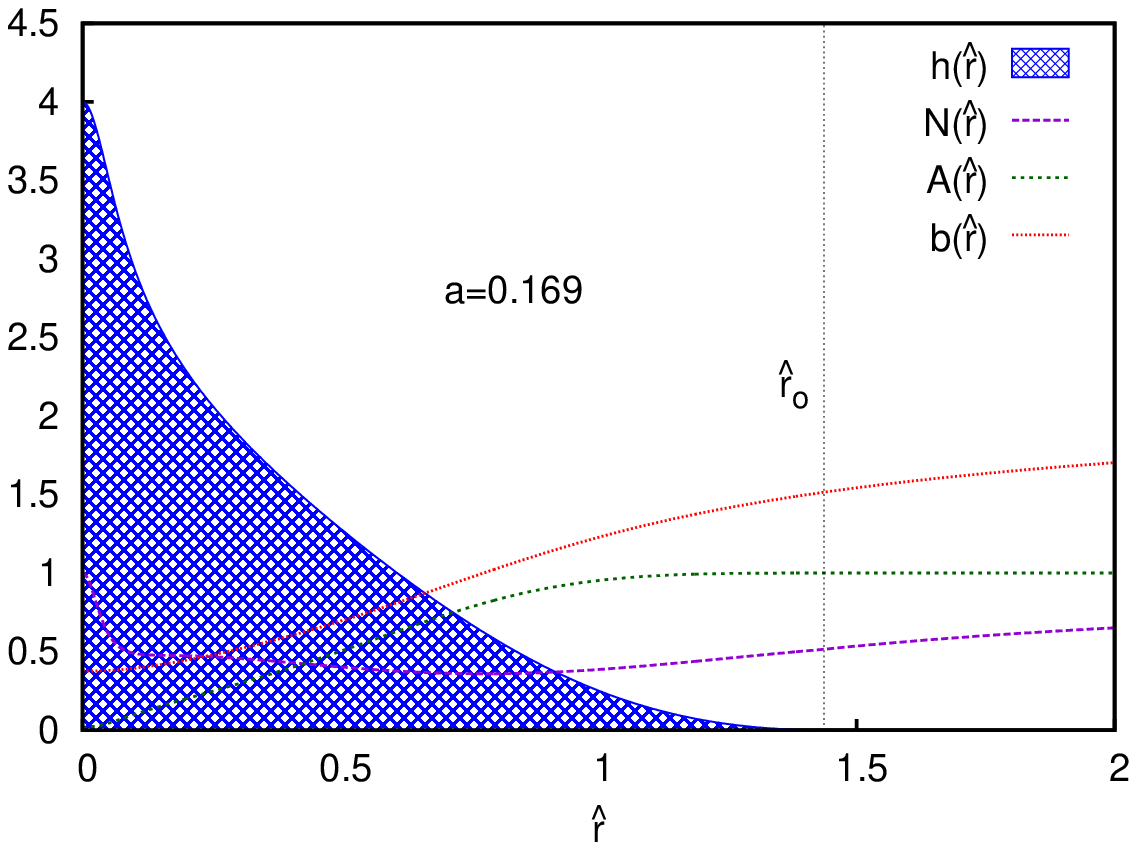}\label{f8c}}
		  \subfigure[][]{\includegraphics[scale=0.68]{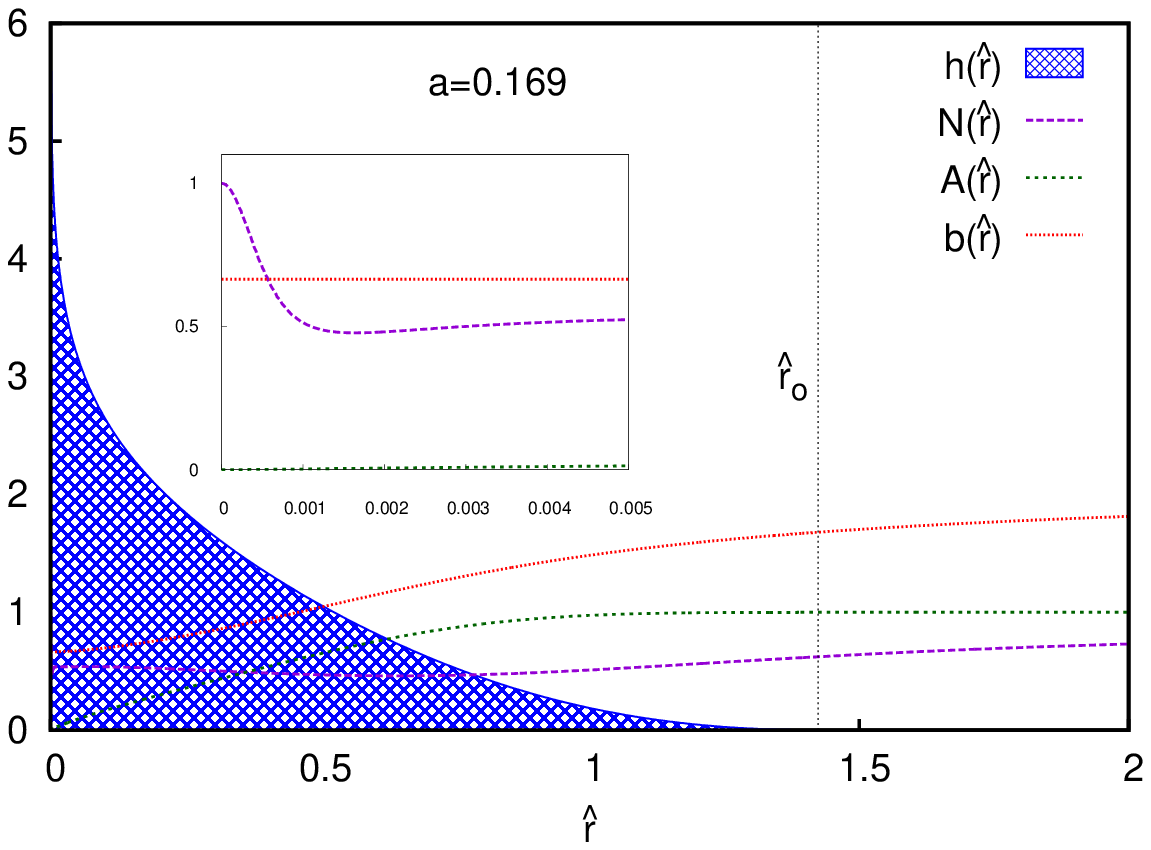}\label{f8d}}}
	\mbox{\subfigure[][]{\includegraphics[scale=0.68]{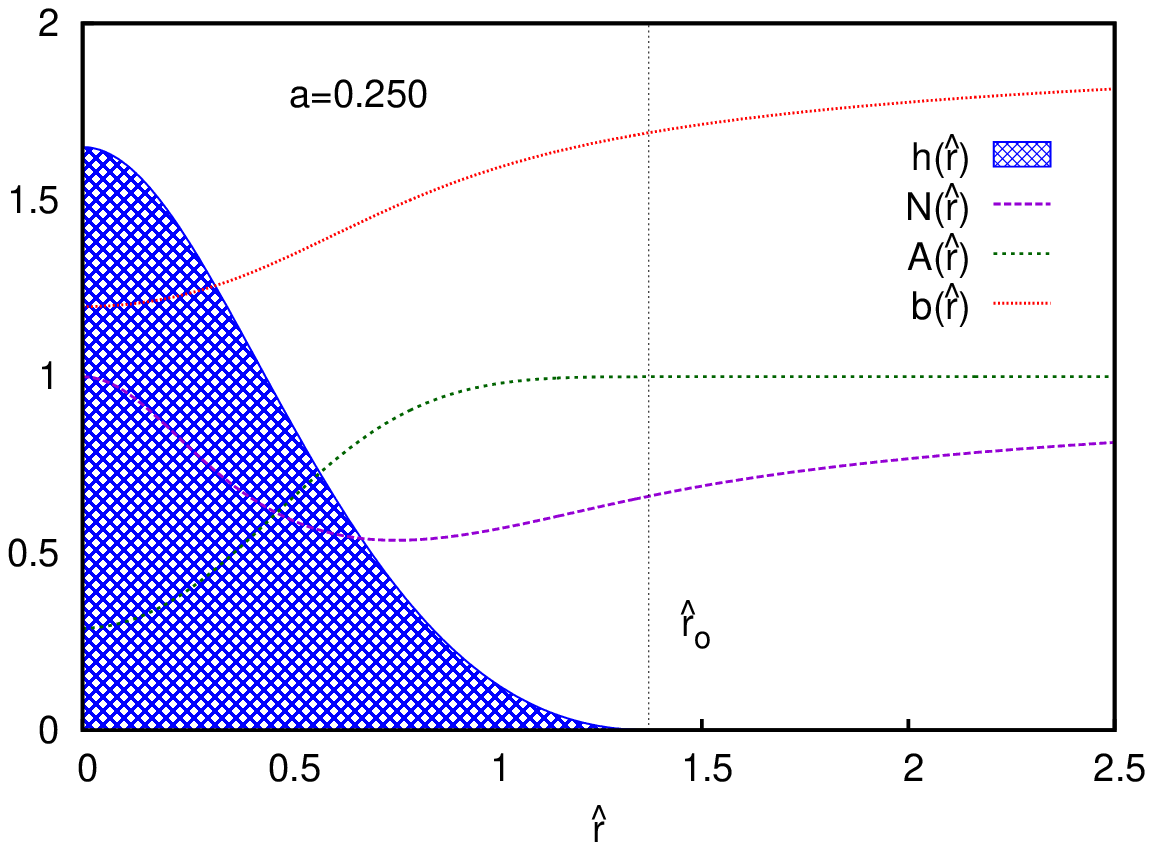}\label{f8e}}
		  \subfigure[][]{\includegraphics[scale=0.68]{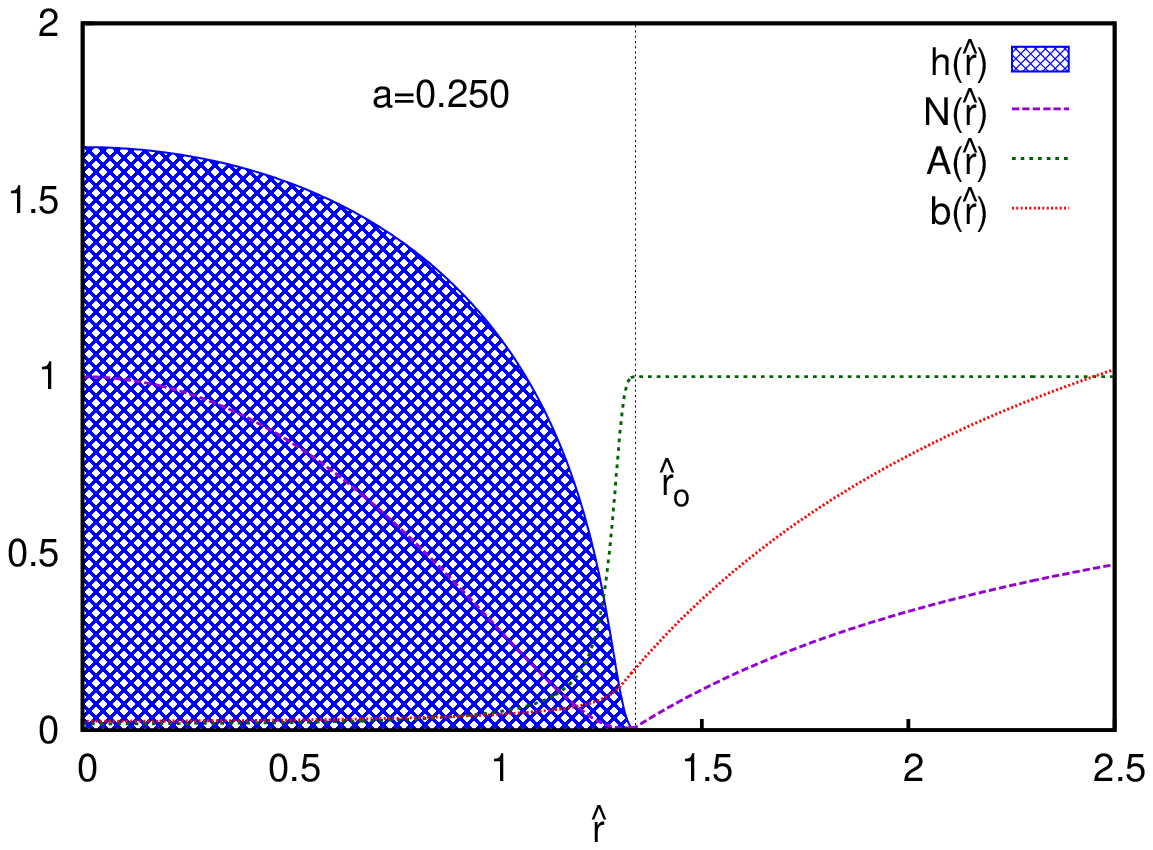}\label{f8f}}}
	\caption{Figs.~(a)-(f) describe the distribution of the four field variables $h(\hat{r})$, $N(\hat{r})$, $A(\hat{r})$ and $b(\hat{r})$ versus the dimensionless radial coordinate $~\hat{r}$), for the ranges shown (which are solutions of the set of four coupled nonlinear differential Eqs.~(16)-(19)). Figs.~(a)-(d) here correspond to the representative points denoted by $P_{1}$ (IA), $P_{2}$ (IA transition to shells), $P_{3}$ (IB2) and $P_{4}$ (IB3), which lie on the phase trajectory corresponding to $a = 0.169$, and Figs.~(e)-(f) correspond to the representative points denoted by $P_{5}$ (IIA) and $P_{6}$ (IIB), which lie on the phase trajectory corresponding to $a = 0.25$.  \label{fig8}}
\end{center}
\end{figure*}

\section{Numerical Solutions}\label{sec:ns}
We now consider the numerical solutions of the coupled nonlinear differential 
equations, Eqs.~(\ref{eq_N})-(\ref{eq_b}), 
subject to the above set of boundary conditions: Eqs.~(\ref{Aro})-(\ref{bcstar}) (cf.  
%obtained by %employing the Newton-Raphson scheme 
%following 
Refs.~\cite{Kleihaus:2009kr,Kleihaus:2010ep,Hartmann:2012da,Hartmann:2013kna,Kumar:2014kna,Kumar:2015sia,Kumar:2016oop,Kumar:2016sxx,Kumar:2017zms}).

Going into the details, the solution of the equations is known outside the outer radius of the boson stars, since there we have the electo-vac equations
with the known Reissner-Nordstr\"om solution. 
In the inner region the scalar field is non-trivial
and vanishes only on the boundary.
In this inner region the equations are then solved 
using the method of shooting, 
subject to the specified boundary conditions, 
compatible with the exterior Reissner-Nordstr\"om solution.

The shooting method is based on converting the two-point boundary value problems into an equivalent initial value problem and this method uses the methods used in solving initial value problems.  The solution is done by assuming initial values at one boundary for the variables that would have been given if the ordinary differential equation were a initial value problem.  The boundary values obtained at the other boundary  is then compared with the actual boundary values. Using some intuitive approach, one tries to get as close to the boundary value as possible.  
In general, one finds discrepancies from the desired boundary values at the other boundary. Now we have a multidimensional root finding problem and we use a root finding method to find the adjustment of the assumed initial values at the starting point that makes the discrepancies at the other boundary vanish.

To solve the coupled nonlinear differential  equations, Eqs.~(\ref{eq_N})-(\ref{eq_b}),  subject to the above set of boundary conditions, Eqs.~(\ref{Aro})-(\ref{bcstar}), we require initial values for the dependent variables $A(0)$, $b(0)$ and $h(0)$ at the inner boundary i.e. at $x=0$. We have to guess initial values in a way such that the boundary conditions at the outer boundary i.e. at $x=1$ are satisfied.

For $A(0)$ we can choose an initial guess between $0$ and $1$ because $A'(r)$ has a positive slope given by eq. (\ref{eq_A}) and $A(r)$ has an upper bound given by $A(r_o)=1$. For $h(0)$ and $b(0)$ we can choose any finite value as an initial guess. For the parameter $a=0$ we can for example, start with an initial guess of $A(0)=1$,  $h(0)=0$ and $b(0)>0$ and the obtained solution is used as an initial guess for the next solution. 
   
In the shooting method we choose values for all the dependent variables, 
consistent with other boundary conditions, at the inner boundary, 
i.e. at $x=0$, and then integrate the initial value problem 
by the Runge-Kutta method of order 4 with an adaptive step size,
where the step size is controlled by the estimate of errors at each step,
until we arrive at the outer boundary, i.e. at $x=1$.

At each iteration step we use the Newton-Raphson method 
to find the zeros of the function representing the errors 
by which the solutions to the initial value problem fails 
to satisfy the boundary condition at $x=1$. 
The set of coupled equations is solved iteratively,
until the prescribed accuracy
and thus the stopping criterion is reached.
We have been using this method here, requiring an accuracy of 6-7 digits.
For further information on the shooting method for boundary value problems
see \cite{Press:1992zz}.

Employing this method,
we determine the domain of existence of the solutions of the above equations by varying the parameter $~a$, the {\it single} remaining parameter of the theory after the redefinitions and rescalings of Sec. III.

In the following sections, we will study the 2D and 3D phase diagrams, the mass $~M$ and charge $~Q$ of these gravitating objects, and subsequently we will study the distribution of the fields with respect to the dimensionless radial coordinate at some representative points on phase trajectories in the phase diagrams of the theory.  

\section{2D Phase Diagrams}\label{sec:2d}
Studies of the 2D phase diagrams of the theory considered in the present work were first carried out  by Kleihaus, Kunz, L\"ammerzahl and List in \cite{Kleihaus:2009kr,Kleihaus:2010ep}, where a bifurcation point in the relevant phase diagrams of the theory was obtained.These studies were extended in our recent work \cite{Kumar:2017zms}, where additional bifurcation points in the relevant 2D phase diagrams theory were found. 

In this section we focus mainly on the additional bifurcation points in the 2D phase diagrams of the theory. In particular, we not only review our recent work \cite{Kumar:2017zms}, which investigates the 2D phase diagrams, but we now study them {\it for larger ranges} of the variables, as depicted in the various results, as compared to our earlier studies \cite{Kumar:2017zms}. Secondly, we also study {\it some additional phase diagrams} of the theory, which were not studied in our earlier work \cite{Kumar:2017zms}, and thirdly, we discuss the next new bifurcation point and present a formula, indicating the presence of an infinite sequence of such bifurcation points.

For this, we first study and discuss the phase diagrams for boson stars as shown in Figs.~\ref{f1a}, \ref{f1b} and \ref{f1c}.  In Fig.~\ref{f1a}, we consider the functions $h(0)$ and $b(0)$, in Fig.~\ref{f1b} the functions $A(0)$ and $b(0)$, and in Fig. \ref{f1c} the functions $A(0)$ and $h(0)$ (where $h(0)$, $b(0)$ and $A(0)$ denote the values of these functions at the centre of the star). Also, the figures shown in the insets in Fig.~\ref{f1a} to \ref{f1c} exhibit particularly interesting areas of these figures with better precision. The asterisks shown in Fig.~\ref{fig1}, corresponding to $h(0)=0$, represent the transition points from the boson stars to boson shells. 

Let us now address the phase diagrams based on the values of the functions at the origin of the boson star, comprising the vector field function $b(0)$ and the scalar field function $h(0)$, that are obtained by studying the solutions for a sequence of values of the parameter $a$. Here we observe very interesting phenomena near specific values of $a$, where the system is seen to have bifurcation points, $B_1\,,\,B_2$ and $B_3$. These bifurcations arise at the following values of $a$,  $a_{c_1}\simeq0.198926, ~a_{c_2}\simeq0.169311$  and $a_{c_3}\simeq0.168308 $, respectively (and a possibility about the existance of further bifurcation points can not be ruled out in principle, if one could somehow increase the numerical accuracy beyond the one already achived by us in the present work (however, practically it is a rather very challenging problem )). %as we have already arrived with six digit results for our bifurcatiobn points)).

Thus, clearly the theory possesses rich physics in the domain $a = 0.22$ to $a \simeq + 0.16\,$.

For a better understanding, let us now divide the phase diagram in the vicinity of the first bifurcation point $B_1$ into four regions, which we denote by IA, IB, IIA and IIB (as depicted in Fig.~\ref{f1a}). The asterisks in Fig.~\ref{f1a} reside on the axis $b(0)$, corresponding to $h(0)=0$, and they mark the transition points from the boson stars to boson shells \cite{Kumar:2017zms}.

First we note that the regions IA, IIA and IIB do not possess any further bifurcation points, while the region IB does contain several more bifurcation points, which indicates that it contains rich physics.
Consequently, we divide the region IB further into the regions IB1, IB2 and IB3 in the vicinity of the first additional bifurcation point $B_2$. 

The region IB3 again contains a further bifucation point, which we denote $B_3$. This suggests a further subdivision of the phase diagram in the vicinity of $B_3$ into the regions IB3a, IB3b and IB3c, as seen in the inset of Fig.~\ref{f1a}. Interestingly, the region IB3b contains closed loops, and the phase diagram in this region resembles the one of the region IB2. 

From these figures we can conclude, that as the value of $a$ changes from $a=0.25 $ to $a=0$,
a lot of new rich physics is encountered.  When varying $a$ from $a=0.25 $ to the critical value $a=a_{c_1}$, the solutions exist in the two distinct domains, IIA and IIB (as depicted in Fig.~\ref{f1a}). But when we decrease $a$ below this critical value $a =a_{c_1}$, the solutions reside in the regions IA and IB. Let us mention here, that for values of $a$ larger than $a=0.25$, the physics remains qualitatively the same as for the value $a=0.25\,$.

When the value of $a$ is decreased
from the first critical value $a=a_{c_1}$ 
to the second critical value $a=a_{c_2}$, 
the curves in the region IA of the phase diagram
we notice that the region IA in the phase diagram 
show a continuous deformation, leading to
the region IB, which exhibits its own rich physics 
as explained above. 

Decreasing the value of $a$ further below $a_{c_2}$, 
we see on the one hand in the region IA a continuous deformation of the curves towards $a=0$. 
However, on the other hand, a third bifurcation point
is encountered in the region IB, dividing this region into IB1, IB2 and IB3. Again, there is a continous deformation of the curves in the region IB1, while the region IB2 again
contains curves forming closed loops. Subdividing the region IB3 into the regions IB3a, IB3b and IB3c, there is again a continuous deformation of the curves in region IB3a, whereas the region IB3b again exhibits closed loops. 

The regularity of the above described phenomenon makes it tempting to conjecture, 
that in fact a whole sequence of further bifurcation points 
may be waiting to be discovered, which might display a self-similar pattern. 
To support this conjecture, we have considered the set of bifurcation points
$a_{c_1}$, $a_{c_2}$ and $a_{c_3}$, and contemplated how to extend this sequence 
to the next bifurcation point $a_{c_4}$. Since the difference between the
bifurcation points $\Delta_n=a_{c_n}-a_{c_{n+1}}$ decreases strongly,
an exponential approach toward a limiting value $a_{c_\infty}$ is expected.
Thus an ansatz for the $n$-th bifurcation point $a_{c_n}$ of the type
\begin{equation}
a_{c_n} = a_{c_\infty} + f a^n
\end{equation}
seems promising. Using the first three bifurcation points we have thus made
a prediction for the fourth one and, indeed, found the fourth bifurcation point
$a_{c_4}=0.16827861$, stretching our numerical calculations and accuracy 
to their current limits.

Since the numerical calculations become increasingly challenging, as one 
progresses from bifurcation point to bifurcation point, we have not been able 
to continue our quest for the next bifurcation points further,
to support this fascinating expectation with even more data. 
But the fact, that our prediction was borne out, does lend considerable
support to our conjecture.
Indeed, a constantly increasing numerical accuracy is required to map 
out the domain of existence of the solutions. 
For instance, to analyze the bifurcation point B3, the value of $a$ needed already 
a specification of 6 decimal digits. The analysis of bifurcation point
B4 needed two more digits. The expontial convergence of the values of the
bifurcation points towards their limiting value, clearly requires a
rapidly increasing number of accucarate digits of the calculations.
Consequently, the global accuracy of the numerical scheme employed represents 
currently the barrier to continue these investigations of potential 
further bifurcation points.

In Fig.~\ref{f1d} we exhibit the radius $\hat{r}_o$ of the star versus the value of the vector field function at the center of the star $b(0)$. As in the previous figures, the region around the first bifurcation point $B_1$ can be divided into the four regions IA, IB and IIA, IIB, and the asterisks represent the transition points from the boson stars to boson shells. However, the oscillations (with respect to $b(0)$) seen in Figs.~\ref{f1a} and \ref{f1b}
in the regions IIA and IB, have transmuted in Fig.~\ref{f1d} into a spiralling behavior. The insets of Fig.~\ref{f1d} magnify parts of the physically interesting region IB. 

\section{3D Phase Diagrams}\label{sec:3d}
In this section, we present a study of the 3D phase diagrams of the charged compact boson stars. Figs.~\ref{f2a} to \ref{f2f} show the 3D diagrams (with different viewing angles) of the scalar field function $h$ and the gauge field function $b$ versus the  metric field function $A$ at the centre of the boson star for a large set of values of the constant $a$. The viewing angles are denoted by $(\theta, \phi)$, where we use the convention that $\theta$ denotes the angle of rotation about the $ox$-axis in the anticlockwise direction and takes values between $~0$ to $\pi$, and that $\phi$ is the angle of rotation about the $oz^\prime$-axis also in the anticlockwise direction and takes values between $~0$ to $2\pi$. Figs.~(a) to (f)  correspond, respectively, to the values $(\theta,\phi)\equiv(60,60),\ (60,175),\ (50,245),\ (135,35),\ (130,70)$ \mbox{ and } $(135,340)$.

Figs.~\ref{f3a} to \ref{f3f} show the 3D diagrams of the scalar field function $h(0)$ and the gauge field function $b(0)$ versus the radius of the star $\hat r_o$ for a large set of values of the constant $~a$.   Figs.~(a) to (f)  correspond, respectively, to the values of the viewing angles $(\theta,\phi)\equiv(60,100),\ (115,15),\ (60,200),$ $\ (40,240),\ (140,60)$ \mbox{ and } $(130,40)$.

\section{Mass $M$ and Charge $Q$}\label{sec:mq}

Let us next consider the global properties of the solutions, their mass $M$ 
%(with $ M := m~\alpha^2 $ 
and their charge $Q$. We present the mass $M$ versus the radius $\hat{r}_{o}$ in Fig.~\ref{f4a}, and the corresponding magnified region of the bifurcations is shown in Fig.~\ref{f4b}. The charge $Q$ versus the radius $\hat{r}_{o}$ is depicted in Figs.~\ref{f5a} and \ref{f5b}, and it is clearly seen to possess a very similar dependence as the mass. 
%This is illustrated in Figs.~\ref{f5a} and  \ref{f5b}.

The stability of the boson stars can be investigated by 
considering the mass $M$ versus the charge $Q$, as shown in Fig.~\ref{fig6}. Additionally it is instructive to consider the mass per unit charge $M/Q$ versus the charge, as shown in Fig.~\ref{fig7}. In Fig.~\ref{f6a} we see that the curves of $M$ versus $Q$, that are located in region IA and concern the smaller values of $a$, all exhibit a monotonic increase from $M=Q=0$ towards their transition points, where the boson shells are formed, and which are marked by crosses. The boson stars forming these curves represent the fundamental boson star solutions (for the respective value of $a$). Therefore these boson stars should be stable and, in fact, the boson stars associated with all curves in region IA should be stable, since these solutions possess 
the lowest mass for a given charge (and parameter $~a$). 

However, it is seen that above a certain value of $~a$, these curves no longer reach a boson shell, but instead their upper endpoint represents a solution, where a throat is formed. In that case the exterior space-time $~r>r_0$ corresponds to the exterior space-time of an extremal RN black hole. As was discussed in great detail in 
\cite{Kleihaus:2009kr,Kleihaus:2010ep}, this happens whenever the value $b(0)=0$ is encountered. Interestingly, 
the curves residing in region IIB encounter solutions with throats at both their endpoints, since both times the value $b(0)=0$ is reached. Because these solutions also possess the lowest mass for a given charge, they are also expected to be stable.

In contrast, in region IIA the solutions exhibit the typical oscillating/spiralling behavior, that is known to arise also for non-compact boson stars. Considered 
in a mass versus charge diagram,  this behavior translates into a sequence of spikes, as seen in the insets of Figs.~\ref{f6a} and \ref{f7a}. In this case the solutions should only be stable along their fundamental branch, which ends when a first spike is encountered at a maximal value of the mass and the charge. Whenever a new spike is encountered, a new unstable mode should arise, analogously to the case of non-compact boson stars.

Addressing now the solutions in the bifurcation regions of the phase diagram, let us consider
the region IB, and starting with the boson stars constituting the limiting curves. Adopting the value $a_{c_1}$,  the two branches of solutions, which limit the region IA, possess lower masses than the two branches of solutions, that limit the region IB. Therefore the lower mass solutions are expected to be more stable.

Indeed, the two branches of solutions, which limit the region IB, might be classically stable, as well, at least until the first extrema of the mass and the charge are reached. From a quantum mechanical point of view, however, they would still be unstable, since the phenomenon of tunnelling might arise. Clearly, beyond these extrema, unstable modes should be present, and the solutions should also be classically unstable.

One can also extend these arguments to all the solutions in region IB. Quantum mechanically they should be unstable, since solutions with lower mass but the same value of the charge reside in the region IA (for all of them). But from a purely classical point of view
the solutions with the lowest mass for a given $a$ within the region IB could still be stable.
In contrast, the higher mass solutions should possess unstable modes and thus be classically unstable.

Figs.~\ref{f6b} and \ref{f6c} depict the mass $M$ versus the charge $Q$ for the respective ranges of $a$  shown (where in Fig.~\ref{f6c} a part of Fig.~\ref{f6b} has been magnified).
Similarly, Figs.~\ref{f7b} and \ref{f7c} depict the mass to charge ratio $M/Q$ versus the charge $Q$ for the respective ranges of $a$ (where again in Fig.~\ref{f7c} a part of Fig.~\ref{f7b} has been magnified). These figures, in fact, illustrate that the solutions in the bifurcation region indeed correspond to higher mass solutions.
%\section{Distribution of Fields with Radial Parameter at some Representative Points of Sample Phase Trajectories}
\section{Distribution of Fields}\label{sec:df}
Let us now consider the distribution of the fields of a representative set of solutions in order to understand how the physical configurations change, as we move along the phase trajectories in the phase digagrams. For that purpose we consider representative points $~P(\hat{r}_o, h(0), b(0), A(0))$, located in the various regions of the phase diagrams. In particular, we choose the following set corresponding to {\it two} different sample values of $~a$, namely for (i) $~a = 0.169 $ and for (ii) $~a = 0.25 $. 

For the case (i) we consider {\it four} particular representative points denoted by $P_{1}, P_{2}, P_{3}, P_{4}$, defined as follows: \\
($P_{1}$): lying in region IA:
\begin{eqnarray}
P &\equiv& ~P(\hat{r}_o, A(0), h(0), b(0)) \nonumber \\
&\equiv& P_1(1.81377, 0.457674, 1.40503,	1. 00) \ \  \nonumber
\end{eqnarray}
($P_{2}$): lying in region IA (corresponding to the transition point to shells)
\begin{eqnarray}
P &\equiv& ~P(\hat{r}_o, A(0), h(0), b(0)) \nonumber \\
&\equiv& P_2(3.14508,	0.0721348, 0.0, 0.0940035)  \ \ \nonumber
\end{eqnarray}
($P_{3}$): lying in region IB2:
\begin{eqnarray} 
P &\equiv& ~P(\hat{r}_o, A(0), h(0), b(0)) \nonumber \\
&\equiv& P_3(1.43695,	0.0186804, 4.0, 0.376553) \ \  \nonumber
\end{eqnarray}
($P_{4}$): lying in region IB3: 
\begin{eqnarray}
P &\equiv& ~P(\hat{r}_o, A(0), h(0), b(0)) \nonumber \\
&\equiv& P_4(1.42383, 0.000677104, 6.0,	0.664592) \ \  \nonumber
\end{eqnarray}
The results for these four representative points are shown in Figs.\ref{f8a}-\ref{f8d}. 

For the case (ii) we consider {\it two} representative points denoted by $P_{5}$ and $P_{6}$, defined as follows:\\
($P_{5}$): lying in region IIA:
\begin{eqnarray}
P &\equiv& ~P(\hat{r}_o, A(0), h(0), b(0) ) \nonumber \\
&\equiv& P_5(1.3712,	0.288522,	1.65, 1.19841)  \ \  \nonumber
\end{eqnarray}
($P_{6}$): lying in region IIB:
\begin{eqnarray}
P &\equiv& ~P(\hat{r}_o, A(0), h(0), b(0) ) \nonumber \\
&\equiv& P_6(1.33932, 0.0180412, 1.65, 0.0253849) \ \  \nonumber
\end{eqnarray}
The results for these are shown in Figs. \ref{f8e}-\ref{f8f}. 

Figs. \ref{f8a}-\ref{f8f} describe the distribution of the four field functions $A(\hat{r})$, $h(\hat{r})$, $b(\hat{r})$ and $N(\hat{r})$) with respect to the dimensionless radial coordinate $~\hat{r}$, which are the solutions of the set of four coupled nonlinear differential Eqs.~(\ref{eq_N})-(\ref{eq_b}). Thus these figures show the values of these fields from the center of the star all the way not only up to the boundary of the compact boson star (where the scalar field $h(\hat{r})$ vanishes), but also beyond the boundary of the star into the exterior region. The electromagnetic field represented by $b(\hat{r})$ and the gravitational field represented by the metric functions $A(\hat{r})$ and $N(\hat{r})$ correspond to long range forces. 

Fig. \ref{f8a} represents a typical compact boson star configuration, located in the phase diagram in region IA. The matter distribution function $h(\hat{r})$ is maximal at the center and decreases monotonically towards the outer boundary $\hat{r}_o$ of the star, where it vanishes. The metric function $N(\hat{r})$ first decreases from its central value $N(0)=1$, necessary for regularity, then it reaches a minimum and increases again towards its asymptotic value of one, necessary for asymptotic flatness. The metric function $A(\hat{r})$ starts from a finite value at the center and increases monotonically towards the boundary of the star, reaching its asymptotic value of one already at the boundary. The gauge field function $b(\hat{r})$, finally, also exhibits a monotonic increase from the center of the star towards a finite value at infinity. Recall, that the spacetime exterior to the star corresponds to a part of a Reissner-Nordstr\"om spacetime.

As we move along the phase trajectory, keeping $a$ fixed, and changing $h(0)$ smoothly, first increasing and the decreasing it, the configurations change smoothly, until $h(0)=0$ is reached. This configuration is shown in Fig. \ref{f8b} and also located in the phase diagram in region IA. It corresponds to the transition point to boson shells. Therefore the matter has been depleted in the interior of the star, so that the matter function vanishes at the center, and then increases slowly towards the outer regions of the star, where it reaches a maximum. From this maximum the matter density then rapdidly drops to zero at the boundary of the star. At the same time, the metric function $N(\hat{r})$ has developed a deep minimum ($N(\hat{r}_{\rm min}) = 0.00811$) close to the outer boundary $\hat{r}_o$ of the star ($\hat{r}_{\rm min}=3.13054 < \hat{r}_o= 3.14508 $), while the metric function $~A(\hat{r})$ shows a steep rise close to the boundary. The gauge field function $b(\hat{r})$ is almost constant inside the star.

The phase trajectory for this value of $a$ consists of several disconnected parts, associated with several of the distinct regions of the phase diagram. Fig. \ref{f8c} corresponds to a point in region IB2, i.e., in the region of the second bifucation point. Here the matter field $h(\hat{r})$ is concentrated much stronger at the center of the star, as compared to a typical star (as seen, e.g., in Fig. \ref{f8a}). The matter density then drops monotonically and rapidly towards the boundary of the star. The center remains regular, as seen from the behavior of the metric functions. 

Fig. \ref{f8d} illustrates the configurations in the region IB3 of the phase diagram, associated with the third bifurcation point. Now the matter is even more strongly localized at the center, dropping off even more steeply towards the boundary of the star. The strong localization of the matter at the center is reflected in the behavior of the metric functions, where the function $N(\hat{r})$ now also exhibits a strong decrease at the center (see inset to verify the regularity condition at the center), while the function $A(\hat{r})$ tends towards zero at the center (but remains finite with $A(0)=0.000677104$).

We conjecture, that for the further bifurcations points, the matter will be even more strongly localized at the center, and the metric function $N(\hat{r})$ will decrease even more strongly as well, while the metric function $A(\hat{r})$ will tend even closer towards zero at the center (but remain finite).

We now turn to another value of $a$, and thus another phase trajectory, which also has disconneted parts.
Fig. \ref{f8e} belongs to a boson star located in region IIA in the phase space diagram. This boson star looks very much like the one in Fig. \ref{f8a}. The maximum at the center has not yet increased much. However, it will increase strongly further along the phase trajectory, as soon as the oscillations will be encountered (which start at $h(0)=1.71$, $b(0)= 1.9349$).

In Fig. \ref{f8f}, finally, a boson star belonging to region IIB is illustrated. This boson star is located on the rather short second part of the phase trajectory for this value of $a$. At both endpoints of this part of the trajectory a special solution is encountered, which is joining to an exterior extremal Reissner-Nordstr\"om solution, where $b(0)=0$ and $N(\hat{r}_o)=0$. The boson star in this figure shows already clear signs of being close to such a special configuration. The metric function $N(\hat{r})$ has already a deep minimum ($N(\hat{r}_{\rm min}) = 0.00603$) close to the boundary of the star ($\hat{r}_{\rm min}= 1.30994 < \hat{r}_o= 3.33932 $), and the gauge field function $b(\hat{r})$ is already close to zero at the center of the star.
\section{Summary and Conclusions}\label{sec:sc}
In this work, we have presented our studies of charged compact boson stars in a theory involving a 
{\it massless} complex scalar field with a conical potential coupled to a U(1) gauge field and gravity. Our presented work focuses mainly on the studies of the phase diagrams of this theory. 

A study of the 2D phase diagrams of this theory was initially undertaken by Kleihaus, Kunz, L\"ammerzahl and List in Refs.~\cite{Kleihaus:2009kr,Kleihaus:2010ep}, where a bifurcation point in the relevant phase diagrams of the theory was obtained. These studies were extended recently in Ref.~\cite{Kumar:2017zms}, where some additional bifurcation points were obtained in the 2D phase diagrams \cite{Kumar:2017zms}.

In this work, we have further extended our earlier studies of the 2D phase diagrams of the theory in two respects.
Firstly we have considered a larger range of the variables involved as compared to our earlier studies, and secondly we have studied additional 2D phase diagrams. 

In addition, we have further extended our above studies to present 3D phase diagrams of the charged compact boson stars. These 3D phase diagrams have been investigated, and a detailed discussion of the various regions of the phase diagrams with respect to the bifurcation points has been presented. The theory is seen to contain rich physics in a particular domain of the phase diagrams, namely in the domain $a=0.25$ to $a\simeq0.16$, where we have identified four bifurcation points $B_1,~B_2,~B_3$ and $B_4$. 
In particular, we have predicted the fourth bifurcation point by analyzing
the first three bifucation points and supposing and exponential
law for the $n$-th bifurcation point, and then verifying its predicted value
by direct computation.
The existence of such a whole sequence of further bifurcation points 
following the exponential law seems therefore a well-supported conjecture, 
but it will remain numerically challenging to construct even further
bifurcation points.

Physical properties of the solutions, including their mass, charge and radius have been investigated. The mass versus charge, and the mass per unit charge versus the charge, have been considered and the arguments concerning the stability of the solutions have been presented.

For any value of $a$, a fundamental branch of compact boson star solutions exists, which is expected to be stable, since the solutions on this branch represent the solutions with the lowest mass for a given charge. In that sense these solutions can be considered to represent the ground state. On the other hand, in the region of the bifurcations additional branches of solutions are present, which possess higher masses for a given charge. These solutions thus correspond to the excited states of the system. The lowest of these might be classically stable and unstable only quantum mechanically. To definitely answer this question, a mode stability analysis needs to be performed. However, such an analysis is beyond the scope of our present work and represents an interesting topic for future investigations.

In addition to this, we have also studied the distribution of field functions $A(\hat{r})$, $h(\hat{r})$, $b(\hat{r})$ and $N(\hat{r})$) versus the dimensionless radial coordinate $\hat{r}$ at some chosen representative points of some specific phase trajectories in the phase diagrams, in order to better understand the physical properties of the boson star configurations in the various parts of the phase diagram. This procedure can, in principle, be used to investigate the distribution of fields at any other representative points on any of the phase trajectories in the phase diagrams for any fixed value of $~a$.

\begin{acknowledgements}\label{sec:ak}
This work was supported in part by the 
DFG Research Training Group 1620 {\sl Models of Gravity} 
as well as by and the COST Action CA16104 {\sl GWverse}.
SK would like to thank the CSIR, New Delhi, for the award 
of a Research Associateship during which a part of this work was carried out. 
\end{acknowledgements}


\begin{thebibliography}{9}
%\cite{Lee:1991ax}
\bibitem{Lee:1991ax}
  T.~D.~Lee, Y.~Pang,
  %``Nontopological solitons,''
  Phys.\ Rept.\  {\bf 221}, 251 (1992).

%\cite{Jetzer:1991jr}
\bibitem{Jetzer:1991jr}
  P.~Jetzer,
  %``Boson stars,''
  Phys.\ Rept.\  {\bf 220}, 163 (1992).

%\cite{Mielke:2000mh}
\bibitem{Mielke:2000mh}
  E.~W.~Mielke, F.~E.~Schunck,
  %``Boson stars: Alternatives to primordial black holes?,''
  Nucl.\ Phys.\  {\bf B564}, 185 (2000).
  %[gr-qc/0001061].

%\cite{Schunck:2003kk}
\bibitem{Schunck:2003kk}
  F.~E.~Schunck, E.~W.~Mielke,
  %``General relativistic boson stars,''
  Class.\ Quant.\ Grav.\  {\bf 20}, R301 (2003).
 %[arXiv:0801.0307 [astro-ph]].

%\cite{Liebling:2012fv}
\bibitem{Liebling:2012fv}
  S.~L.~Liebling and C.~Palenzuela,
  ``Dynamical Boson Stars,''
  Living Rev.\ Rel.\  {\bf 15}, 6 (2012).
  doi:10.12942/lrr-2012-6
  [arXiv:1202.5809 [gr-qc]].

%\cite{Kaup:1968zz}
\bibitem{Kaup:1968zz}
  D.~J.~Kaup,
  ``Klein-Gordon Geon,''
  Phys.\ Rev.\  {\bf 172}, 1331 (1968).
  doi:10.1103/PhysRev.172.1331

%\cite{Feinblum:1968nwc}
\bibitem{Feinblum:1968nwc}
  D.~A.~Feinblum and W.~A.~McKinley,
  ``Stable States of a Scalar Particle in Its Own Gravational Field,''
  Phys.\ Rev.\  {\bf 168}, no. 5, 1445 (1968).
  doi:10.1103/PhysRev.168.1445
%\cite{Ruffini:1969qy}
\bibitem{Ruffini:1969qy}
  R.~Ruffini and S.~Bonazzola,
  ``Systems of selfgravitating particles in general relativity and the concept of an equation of state,''
  Phys.\ Rev.\  {\bf 187}, 1767 (1969).
  doi:10.1103/PhysRev.187.1767

%\cite{Colpi:1986ye}
\bibitem{Colpi:1986ye}
  M.~Colpi, S.~L.~Shapiro, I.~Wasserman,
  ``Boson Stars: Gravitational Equilibria of Selfinteracting Scalar Fields,''
  Phys.\ Rev.\ Lett.\  {\bf 57}, 2485 (1986).

%\cite{Friedberg:1986tq}
\bibitem{Friedberg:1986tq}
  R.~Friedberg, T.~D.~Lee, Y.~Pang,
  ``Scalar Soliton Stars And Black Holes,''
  Phys.\ Rev.\  {\bf D35}, 3658 (1987).

%\cite{Meliani:2015zta}
\bibitem{Meliani:2015zta}
  Z.~Meliani, F.~H.~Vincent, P.~Grandcl\'{e}ment, E.~Gourgoulhon, R.~Monceau-Baroux and O.~Straub,
  ``Circular geodesics and thick tori around rotating boson stars,''
  Class.\ Quant.\ Grav.\  {\bf 32}, 235022 (2015)
  doi:10.1088/0264-9381/32/23/235022
  [arXiv:1510.04191 [astro-ph.HE]].

%\cite{Meliani:2017ktw}
\bibitem{Meliani:2017ktw}
  Z. Meliani, F. Casse, P. Grandcl\'{e}ment, E. Gourgoulhon and F. Dauvergne
  ``On tidal disruption of clouds and disk formation near boson stars,''
  Class.\ Quant.\ Grav.\  {\bf 34}, 225003 (2017)
  doi:10.1088/1361-6382/aa8fb5


%\cite{Macedo:2013jja}
\bibitem{Macedo:2013jja} 
  C.~F.~B.~Macedo, P.~Pani, V.~Cardoso and L.~C.~B.~Crispino,
  ``Astrophysical signatures of boson stars: quasinormal modes and inspiral resonances,''
  Phys.\ Rev.\ D {\bf 88}, no. 6, 064046 (2013)
  doi:10.1103/PhysRevD.88.064046
  [arXiv:1307.4812 [gr-qc]].

%\cite{Palenzuela:2017kcg}
\bibitem{Palenzuela:2017kcg} 
  C.~Palenzuela, P.~Pani, M.~Bezares, V.~Cardoso, L.~Lehner and S.~Liebling,
  ``Gravitational Wave Signatures of Highly Compact Boson Star Binaries,''
  Phys.\ Rev.\ D {\bf 96}, no. 10, 104058 (2017)
  doi:10.1103/PhysRevD.96.104058
  [arXiv:1710.09432 [gr-qc]].

%\cite{Hartmann:2012wa}
\bibitem{Hartmann:2012wa} 
  B.~Hartmann and J.~Riedel,
  ``Glueball condensates as holographic duals of supersymmetric Q-balls and boson stars,''
  Phys.\ Rev.\ D {\bf 86}, 104008 (2012)
  doi:10.1103/PhysRevD.86.104008
  [arXiv:1204.6239 [hep-th]].

%\cite{Herdeiro:2014goa}
\bibitem{Herdeiro:2014goa}
  C.~A.~R.~Herdeiro and E.~Radu,
  ``Kerr black holes with scalar hair,''
  Phys.\ Rev.\ Lett.\  {\bf 112}, 221101 (2014)
  doi:10.1103/PhysRevLett.112.221101
  [arXiv:1403.2757 [gr-qc]].

%\cite{Herdeiro:2015tia}
\bibitem{Herdeiro:2015tia}
  C.~A.~R.~Herdeiro, E.~Radu and H.~Runarsson,
  ``Kerr black holes with self-interacting scalar hair: hairier but not heavier,''
  Phys.\ Rev.\ D {\bf 92}, 084059 (2015)
  doi:10.1103/PhysRevD.92.084059
  [arXiv:1509.02923 [gr-qc]].

%\cite{Collodel:2017end}
\bibitem{Collodel:2017end} 
  L.~G.~Collodel, B.~Kleihaus and J.~Kunz,
  ``Static Orbits in Rotating Spacetimes,''
  Phys.\ Rev.\ Lett.\  {\bf 120}, no. 20, 201103 (2018)
  doi:10.1103/PhysRevLett.120.201103
  [arXiv:1711.05191 [gr-qc]].

%\cite{Jetzer:1989av}
\bibitem{Jetzer:1989av}
  P.~Jetzer and J.~J.~van der Bij,
  ``Charged Boson Stars,''
  Phys.\ Lett.\ B {\bf 227}, 341 (1989).
  doi:10.1016/0370-2693(89)90941-6

%\cite{Jetzer:1989us}
\bibitem{Jetzer:1989us}
  P.~Jetzer,
  ``Stability of Charged Boson Stars,''
  Phys.\ Lett.\ B {\bf 231}, 433 (1989).
  doi:10.1016/0370-2693(89)90689-8

%\cite{Jetzer:1993nk}
\bibitem{Jetzer:1993nk}
  P.~Jetzer, P.~Liljenberg and B.~S.~Skagerstam,
  ``Charged boson stars and vacuum instabilities,''
  Astropart.\ Phys.\  {\bf 1}, 429 (1993)
  doi:10.1016/0927-6505(93)90008-2
  [astro-ph/9305014].

%\cite{Pugliese:2013gsa}
\bibitem{Pugliese:2013gsa}
  D.~Pugliese, H.~Quevedo, J.~A.~Rueda H. and R.~Ruffini,
  ``On charged boson stars,''
  Phys.\ Rev.\ D {\bf 88}, 024053 (2013)
  doi:10.1103/PhysRevD.88.024053
  [arXiv:1305.4241 [astro-ph.HE]].

%\cite{Delgado:2016jxq}
\bibitem{Delgado:2016jxq}
  J.~F.~M.~Delgado, C.~A.~R.~Herdeiro, E.~Radu and H.~Runarsson,
  ``Kerr–Newman black holes with scalar hair,''
  Phys.\ Lett.\ B {\bf 761}, 234 (2016)
  doi:10.1016/j.physletb.2016.08.032
  [arXiv:1608.00631 [gr-qc]].
  %%CITATION = doi:10.1016/j.physletb.2016.08.032;%%

%\cite{Collodel:2019ohy}
\bibitem{Collodel:2019ohy} 
  L.~G.~Collodel, B.~Kleihaus and J.~Kunz,
  ``On the Structure of Rotating Charged Boson Stars,''
  arXiv:1901.11522 [gr-qc]

%\cite{Arodz:2008jk}
\bibitem{Arodz:2008jk} 
  H.~Arodz and J.~Lis,
  ``Compact Q-balls in the complex signum-Gordon model,''
  Phys.\ Rev.\ D {\bf 77}, 107702 (2008)
  doi:10.1103/PhysRevD.77.107702
  [arXiv:0803.1566 [hep-th]].

%\cite{Arodz:2008nm}
\bibitem{Arodz:2008nm} 
  H.~Arodz and J.~Lis,
  ``Compact Q-balls and Q-shells in a scalar electrodynamics,''
  Phys.\ Rev.\ D {\bf 79}, 045002 (2009)
  doi:10.1103/PhysRevD.79.045002
  [arXiv:0812.3284 [hep-th]].

%\cite{Kleihaus:2009kr}
\bibitem{Kleihaus:2009kr} 
  B.~Kleihaus, J.~Kunz, C.~Lammerzahl and M.~List,
  ``Charged Boson Stars and Black Holes,''
  Phys.\ Lett.\ B {\bf 675}, 102 (2009)
  doi:10.1016/j.physletb.2009.03.066
  [arXiv:0902.4799 [gr-qc]].

%\cite{Kleihaus:2010ep}
\bibitem{Kleihaus:2010ep} 
  B.~Kleihaus, J.~Kunz, C.~Lammerzahl and M.~List,
  ``Boson Shells Harbouring Charged Black Holes,''
  Phys.\ Rev.\ D {\bf 82}, 104050 (2010)
  doi:10.1103/PhysRevD.82.104050
  [arXiv:1007.1630 [gr-qc]].

%\cite{Hartmann:2012da}
\bibitem{Hartmann:2012da}
  B.~Hartmann, B.~Kleihaus, J.~Kunz and I.~Schaffer,
  ``Compact Boson Stars,''
  Phys.\ Lett.\ B {\bf 714}, 120 (2012)
  doi:10.1016/j.physletb.2012.06.067
  [arXiv:1205.0899 [gr-qc]].

%\cite{Hartmann:2013kna}
\bibitem{Hartmann:2013kna} 
  B.~Hartmann, B.~Kleihaus, J.~Kunz and I.~Schaffer,
  ``Compact (A)dS Boson Stars and Shells,''
  Phys.\ Rev.\ D {\bf 88}, no. 12, 124033 (2013)
  doi:10.1103/PhysRevD.88.124033
  [arXiv:1310.3632 [gr-qc]].

%\cite{Kumar:2014kna}
\bibitem{Kumar:2014kna} 
  S.~Kumar, U.~Kulshreshtha and D.~Shankar Kulshreshtha,
  ``Boson stars in a theory of complex scalar fields coupled to the U(1) gauge field and gravity,''
  Class.\ Quant.\ Grav.\  {\bf 31}, 167001 (2014)
  doi:10.1088/0264-9381/31/16/167001
  [arXiv:1605.07210 [hep-th]].

%\cite{Kumar:2015sia}
\bibitem{Kumar:2015sia} 
  S.~Kumar, U.~Kulshreshtha and D.~S.~Kulshreshtha,
  ``Boson stars in a theory of complex scalar field coupled to gravity,''
  Gen.\ Rel.\ Grav.\  {\bf 47}, no. 7, 76 (2015)
  doi:10.1007/s10714-015-1918-0
  [arXiv:1605.07015 [hep-th]].

%\cite{Kumar:2016oop}
\bibitem{Kumar:2016oop} 
  S.~Kumar, U.~Kulshreshtha and D.~S.~Kulshreshtha,
  ``New Results on Charged Compact Boson Stars,''
  Phys.\ Rev.\ D {\bf 93}, no. 10, 101501 (2016)
  doi:10.1103/PhysRevD.93.101501
  [arXiv:1605.02925 [hep-th]].

%\cite{Kumar:2016sxx}
\bibitem{Kumar:2016sxx} 
  S.~Kumar, U.~Kulshreshtha and D.~S.~Kulshreshtha,
  ``Charged compact boson stars and shells in the presence of a cosmological constant,''
  Phys.\ Rev.\ D {\bf 94}, no. 12, 125023 (2016)
  doi:10.1103/PhysRevD.94.125023
  [arXiv:1709.09449 [hep-th]].

%\cite{Kumar:2017zms}
\bibitem{Kumar:2017zms} 
  S.~Kumar, U.~Kulshreshtha, D.~S.~Kulshreshtha, S.~Kahlen and J.~Kunz,
  ``Some new results on charged compact boson stars,''
  Phys.\ Lett.\ B {\bf 772}, 615 (2017)
  doi:10.1016/j.physletb.2017.07.041
  [arXiv:1709.09445 [hep-th]].

%\cite{Press:1992zz}
\bibitem{Press:1992zz} 
  W.~H.~Press, S.~A.~Teukolsky, W.~T.~Vetterling and B.~P.~Flannery,
  ``Numerical Recipes in FORTRAN: The Art of Scientific Computing'',
  second edition (Cambridge University Press 1996)
\end{thebibliography}
\end{document}